\documentclass[pra,aps,amssymb,twocolumn,noshowpacs]{revtex4-1}
\usepackage{color}
\usepackage{graphicx}
\usepackage{hyperref}
\usepackage{amsmath}
\usepackage{latexsym}
\usepackage{amssymb}
\usepackage{mathrsfs}
\usepackage{mathtools}
\usepackage{layout}
\usepackage{verbatim}
\usepackage{multirow}
\usepackage{bm}
\usepackage{amsfonts,epsfig}

\begin{document}
\title{Rydberg quantum computation with nuclear spins in two-electron neutral atoms}

\date{\today}
\author{Xiao-Feng Shi}
\affiliation{School of Physics and Optoelectronic Engineering, Xidian University, Xi'an 710071, China}

\begin{abstract}
  Alkaline-earth-like~(AEL) atoms with two valence electrons and a nonzero nuclear spin can be excited to Rydberg state for quantum computing. Typical AEL ground states possess no hyperfine splitting, but unfortunately a GHz-scale splitting seems necessary for Rydberg excitation. Though strong magnetic fields can induce a GHz-scale splitting, weak fields are desirable to avoid noise in experiments. Here, we provide two solutions to this outstanding challenge with realistic data of well-studied AEL isotopes. In the first theory, the two nuclear spin qubit states $|0\rangle$ and $|1\rangle$ are excited to Rydberg states $|r\rangle$ with detuning $\Delta$ and 0, respectively, where a MHz-scale detuning $\Delta$ arises from a weak magnetic field on the order of 1~G. With a proper ratio between $\Delta$ and $\Omega$, the qubit state $|1\rangle$ can be fully excited to the Rydberg state while $|0\rangle$ remains there. In the second theory, we show that by choosing appropriate intermediate states a two-photon Rydberg excitation can proceed with only one nuclear spin qubit state. The second theory is applicable whatever the magnitude of the magnetic field is. These theories bring a versatile means for quantum computation by combining the broad applicability of Rydberg blockade and the incomparable advantages of nuclear-spin quantum memory in two-electron neutral atoms.

\end{abstract}
\maketitle

\section{introduction}\label{sec01}

It is widely believed that quantum information processing can carry out fantastic tasks beyond the capability of classical technology~\cite{Nielsen2000,Ladd2010}, although unfortunately a universal quantum computer has not yet been realized especially because of the fast dissipation on the timescale of each system useful for realizing a quantum information processor~\cite{Blatt2008,You2005,You2011,Awschalom2013,Devoret2013}. Recently, neutral atoms have been recognized as a new promising platform for large-scale quantum computing~\cite{PhysRevLett.85.2208,Lukin2001,Saffman2010,Saffman2016,Weiss2017,Adams2019} because quantum entanglement based on the dipole blockade between neutral Rydberg atoms can be generated rapidly in a large qubit array~\cite{Wilk2010,Isenhower2010,Zhang2010,Maller2015,Jau2015,Zeng2017,Levine2018,Picken2018,Levine2019,Graham2019,Jo2019,Madjarov2020}. However, models for reliable and practical quantum computing need entangling gates with a fidelity beyond, for example, $0.999$~\cite{Crow2016}, while the best experimental fidelity of two-qubit entangling gates with alkali-metal atoms is below $0.98$~\cite{Levine2019,Graham2019}. 

A new route toward neutral-atom quantum computation is by using atoms with two valence electrons and a nonzero nuclear spin, i.e., some alkaline-earth-metal or lanthanide atoms, which we call alkaline-earth-like~(AEL) atoms. Compared to alkali-metal atoms, well-studied AEL atoms such as strontium and ytterbium can be more easily cooled to very low temperatures~\cite{Yamamoto2016,Saskin2018,Cooper2018,Covey2019,Madjarov2020} or to the vibrational ground state~\cite{Norcia2018}, their nuclear spin states are insensitive to background magnetic noise and can be preserved in the process of cooling~\cite{Reichenbach2007}, and long-lived trapping of both the ground and Rydberg states is realizable~\cite{wilson2019trapped}. These properties compare favorably to those of alkali-metal atoms concerning quantum control for entanglement generation. A recent Rydberg blockade experiment~\cite{Madjarov2020} with alkaline-earth-metal atoms reported a two-atom entanglement fidelity $99.5\%$, while the best number with alkali-metal atoms is $97.4\%$~\cite{Levine2018,Levine2019}. However, the entanglement in~\cite{Madjarov2020} is between a Rydberg state and a metastable state $(5s5p)^{3}P_0$ of the nuclear-spin-free strontium-88. In principle, by Rydberg blockade one can entangle nuclear spin qubit states in the ground-state space, but its realization faces an outstanding challenge, namely, that there is no hyperfine splitting in the ground state of well-known AEL fermions like strontium-87 and ytterbium-171 and it is unlikely to leave one nuclear spin qubit state unaffected when rotating the other qubit state back and forth to Rydberg state with a megahertz rate.

\begin{figure}
\includegraphics[width=3.4in]
{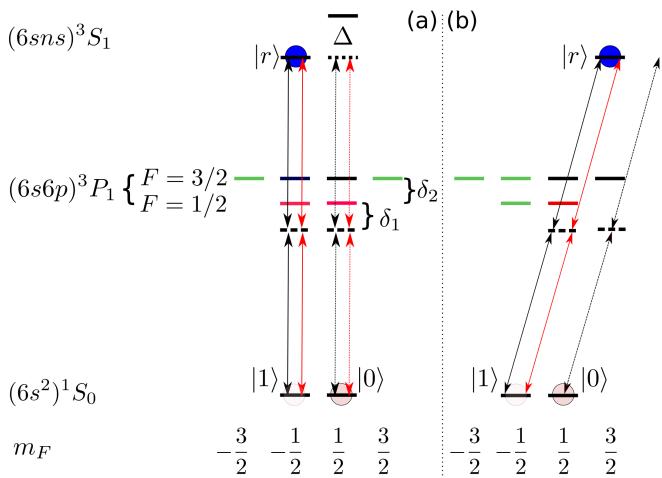}
\caption{ Level diagram with $^{171}$Yb for the two theories of selectively exciting one~($|1\rangle$) of two nuclear spin qubit states $|0(1)\rangle$ to the Rydberg state $|r\rangle$. (a) Theory 1: when a megahertz-scale frequency separation $\Delta$ between the two transitions $|0\rangle\leftrightarrow|r\rangle$ and $|1\rangle\leftrightarrow|r\rangle$ arises from a magnetic field, and when the Rabi frequency $\Omega$ for $|1\rangle\leftrightarrow|r\rangle$ satisfies the condition $\overline{\Omega}=\sqrt{\Omega^2+\Delta^2}=2N\Omega$ with an integer $N$ (if the Rydberg Rabi frequencies on both qubits are equal), a $\pi$ pulse excites $|1\rangle$ to $|r\rangle$, but meanwhile the input state $|0\rangle$ undergoes $N$ cycles of detuned Rabi rotation, returning to itself. (b) Theory 2: a two-photon excitation via the lowest $^3P_1$ state can be tuned resonant to a $^3S_1$ Rydberg state. Due to selection rules, only the nuclear spin state $m_F=-1/2$ is excited to the Rydberg state. The level diagram is not to scale. Theory 2 is applicable even if the mismatch of the transition frequencies for $|0(1)\rangle\leftrightarrow|r\rangle$ is negligible,  \label{figure2} }
\end{figure}

In this work, two theories, Theory 1 and Theory 2, are given to selectively and rapidly transfer one nuclear spin qubit state from the ground state to a Rydberg state of an AEL atom. For Theory 1 shown in Fig.~\ref{figure2}(a), a magnetic field $\mathbf{B}$ on the order of 0.1mT for specifying the quantization axis splits the degeneracy of the two transitions $|0\rangle\leftrightarrow|r\rangle$ and $|1\rangle\leftrightarrow|r\rangle$ with a frequency mismatch $\Delta$ on the order of megahertz. A megahertz-rate Rabi flopping between ground and Rydberg states is required to avoid decoherence in the Rydberg states, so the Rabi frequency $\Omega$ for the rotation $|1\rangle\leftrightarrow|r\rangle$ should be on the order of megahertz, leading to an undesired detuned Rabi oscillation between $|0\rangle$ and $|r\rangle$. The generalized Rabi frequency $\overline{\Omega}$ for the detuned Rabi oscillation of $|0\rangle$ is $\sqrt{\Omega^2+\Delta^2}$~\cite{Shi2017}, and, remarkably, the condition $\overline{\Omega}/\Omega=2N$ can lead to $N$ cycles for the detuned Rabi rotations between $|0\rangle$ and $|r\rangle$ when $|1\rangle$ is excited to $|r\rangle$, where $N$ is an integer. Although there is a phase accumulation in $|0\rangle$, it can be canceled by single-qubit phase gates for designing a controlled-Z~($C_Z$) gate as shown later. For Theory 2 shown in Fig.~\ref{figure2}(b), a two-photon Rydberg excitation $|1\rangle\rightarrow|r\rangle$ is achieved via circularly polarized laser fields. We show by realistic data that it is possible to choose appropriate intermediate states and Rydberg states so that we can completely suppress the Rydberg excitation of $|0\rangle$ when $|1\rangle$ is excited to Rydberg states. Theory 2 does not rely on a MHz-scale $\Delta$ and, in fact, any value of $\Delta$ is fine. Theory 2 can also be executed by pulse-shaping methods such as stimulated Raman adiabatic passage~(STIRAP)~\cite{Bergmann1998,Kral2007,Vitanov2017} that are useful for Rydberg atom quantum entanglement~\cite{PhysRevLett.85.2208,Moller2008,Muller2009,Goerz2011,Beterov2011,Muller2011,Keating2013,Petrosyan2013,Beterov2013,Muller2014,Goerz2014,Beterov2014,Keating2015,Beterov2016jpb,Theis2016,Wu2017,Petrosyan2017,Kang2018,Levine2019,Omran2019,Liao2019,Sun2020,Shi2020jpb,Mitra2020,Beterov2020,Saffman2020,Kang2020,Guo2020,Khazali2020}. In comparison, Theory 1 can not be implemented with STIRAP, but it can be used with one-photon excitation of $p$-orbital Rydberg states~\cite{Hankin2014,Jau2015}.

After reviewing the present stage of AEL-based quantum computing in Sec.~\ref{sec02}, we detail our theories with ytterbium in Sec.~\ref{sec03} and briefly study strontium in Sec.~\ref{sec04} because these two elements are well studied in experiments, as shown in Refs.~\cite{Yamamoto2016,Saskin2018,wilson2019trapped} and~\cite{Madjarov2020,Cooper2018,Covey2019,Norcia2018,Teixeira2020}, respectively. Discussions and conclusions are given in Sec.~\ref{sec05} and~\ref{sec06}, respectively. Although Sec.~\ref{sec03} contains necessary details, more are grouped in the appendixes to avoid blurring the essence of the theories.

\section{Nuclear spin states as qubit states}\label{sec02}
\subsection{ A brief review of previous methods }\label{sec02A}
It will be useful to briefly review previous gate proposals with AEL atoms. The motivation is that they have stable nuclear spin states which can be used to store information, and there can be collisional interaction that enables entanglement~\cite{Jaksch1999,Calarco2000,Hayes2007}. Moreover, for elements with nonzero nuclear spins, an electric singlet-triplet clock transition from the ground state can be decoupled from the nuclear spin. As a consequence, the atoms can be cooled without changing the quantum states of the nuclear spin, which is an awesome benefit to store the information in the state space of the nuclear spin~\cite{Reichenbach2007}. This in principle means that when qubits heat up after many cycles of quantum control as required in quantum computing, they can be cooled again without changing the information stored in the quantum system. Although there is only one ground level $^1S_0$ for well studied AEL atoms, their first excited $^3P_0$ state has a long lifetime $\tau$, which can be used for shelving one of the nuclear spin qubit states for information readout; for the case of ytterbium studied in this work, $\tau$ is about $26~$s~\cite{Covey2019prappl} which can exceed the time scale of relevant quantum control by many times. Besides, coding information in the nuclear spin states can avoid environmental noise such as fluctuation of magnetic fields. These several advantages motivated proposals for entangling atoms with two valence electrons by merging them from different sites for the induction of a blockade or exchange interaction~\cite{Hayes2007,Daley2008}, where the latter has been studied in experiments~\cite{Cappellini2014,Scazza2014,Kaufman2015}. Till now, several theories have been put forward for showing that high-fidelity quantum gates should be feasible via the collisional interaction~\cite{Gorshkov2009,Daley2011,Daley2011qip,Pagano2019,Jensen2019}.

Another method is to use the optical clock transition $^1S_0\leftrightarrow^3P_0$ of AEL neutral atoms~\cite{Stock2008}, where a qubit is defined by the electronic ground and long-lived clock states. This method easily offers frequency-selective excitation to Rydberg states. As in the experiment~\cite{Madjarov2020}, the $^{3}P_0$ clock state was coherently and rapidly excited to a Rydberg state, leading to high-fidelity two-atom entanglement between the Rydberg and optical-clock states.

\subsection{Difficulties in selective Rydberg excitation of one nuclear spin qubit state }
In this work, we show that entanglement between nuclear spins in two-electron atoms can be efficiently generated by Rydberg interactions. Compared to the collisional gates~\cite{Daley2008,Gorshkov2009,Daley2011,Daley2011qip,Pagano2019,Jensen2019}, our Rydberg gate is implemented on a much shorter microsecond timescale. But there are difficulties to achieve our goal.

The first difficulty is that Rydberg states are not stable. Different isotopes have different properties. We take $^{171}$Yb as an example to briefly show the difficulties one by one while more details are grouped in Appendixes~\ref{appendixspinorbit}-\ref{appendixE}. The qubit is defined by two nuclear spin states, $|0(1)\rangle\equiv |(6s^2)^1S_0,~m_I=\pm1/2\rangle$, among which one should be excited to a Rydberg state $|r\rangle\equiv|(6sns)^3S_1,F=I+1\rangle$ for entanglement generation. As estimated in Appendix~\ref{appendixC}, the lifetime of $|r\rangle$ is about $\tau=330~\mu$s for $n=70$ at room temperatures. Rydberg excitation should take place within a time much shorter than $\tau$.

The second difficulty is that for frequently studied AEL atoms, there is no hyperfine splitting in the ground states. When laser fields are sent to qubits for exciting $|1\rangle$ to Rydberg states, the other qubit state $|0\rangle$ can also be excited. Moreover, because the dipole matrix elements between the ground state and a Rydberg state is rather tiny, a two-photon Rydberg excitation is more favorable. A standard way to achieve this is to use a largely detuned intermediate $p$-orbital state $|p\rangle$ for Rydberg excitation. To avoid using ultraviolet light of too short wavelengths, the lowest $^1P_1$ or $^3P_{0,1,2}$ states can be used as the intermediate state. But for the lowest $^1P_1$ state whose excitation from the ground state is spin and dipole allowed, it is difficult to find available laser powers to compensate the fast dissipation of $^1P_1$. For the lowest $^3P_{0,1,2}$ manifold, it is unclear whether the transition can be fast enough with available technology.  

The third difficulty lies in that weak magnetic fields are preferred in experiments when Rydberg states need to be coherently excited. Although strong magnetic fields can in principle lead to a GHz-scale $\Delta$, they are incompatible with Rydberg atom quantum science. First, Rydberg excitation of neutral atoms requires sending laser fields to the vacuum chamber in various directions that cause trouble in setting up devices for generating strong magnetic fields. Second, qubits are often left in free flight during Rydberg excitation so that an inhomogeneous distribution of $\mathbf{B}$ leads to an extra dephasing~\cite{Saffman2005,Saffman2011}, and the gradient of $\mathbf{B}$ along the motional direction of the qubits can be larger when stronger fields are employed. The magnitudes of $\mathbf{B}$ in Rydberg gate experiments by alkali-metal atoms were $9,~11.5,~3.7,~1.5,~4.8,~3,~1.5,~7.5$, $8.5$, and $6$~G in~\cite{Wilk2010,Isenhower2010,Zhang2010,Maller2015,Jau2015,Zeng2017,Levine2018,Picken2018,Levine2019,Graham2019}, respectively. In a recent experiment with AEL qubits, a $710$~G field was used for state initialization~\cite{Madjarov2020}, but $\mathbf{B}$ was switched to $71$~G during Rydberg excitation~\cite{Madjarov2020}.

\section{Selective Rydberg excitation of a nuclear spin state of $^{171}$Yb}\label{sec03}

\subsection{Theory 1}\label{theory1sub} 
  Theory 1 is is shown in Fig.~\ref{figure2}(a) with $\pi$ polarized laser fields where $|1\rangle$ is resonantly excited. The complicated level scheme of AEL requires extra efforts to design the scheme to excite a nuclear spin state. 

For $^{171}$Yb, there is no quadrupole interaction, and the Hamiltonian for the Zeeman interaction in the presence of a magnetic field $\mathbf{B}$ is
\begin{eqnarray}
 \hat{H} &=& A\hat{\mathbf{I}}\cdot \hat{\mathbf{J}}+\mathbf{B} \cdot (g_{ J} \mu_B\hat{\mathbf{J}} - g_{I} \mu_n\hat{\mathbf{I}})\label{hfmagnetic01}
\end{eqnarray}
where $A$ is the nuclear magnetic dipole constant, $\hat{\mathbf{I}}$ and $\hat{\mathbf{J}}$ are the nuclear and electron spin operators, respectively, and $g_{ I(J)}$ is the nuclear~(electron) g factor. Here, $\mu_{B}$ is the Bohr magneton and $\mu_{n}$ is the nuclear magneton, which is equal to $\mu_{n}=0.4919\mu_N$ for $^{171}$Yb~\cite{Porsev2004} and $\mu_N\approx\mu_B/1836$. Because different electron orbits have different contacts with the nucleus, different states have different values of $A$~\cite{PhysRev.178.18,Berends1992,Deilamian1993,Zinkstok2002}. For the ground state the total electron angular momentum is zero, and the only term left in Eq.~(\ref{hfmagnetic01}) is the Zeeman splitting $g_{I} \mu_nB_z$ between $|0\rangle$ and $|1\rangle$. Larger fields can be used for optical pumping during the state initialization, afterward the field can be lowered, as in Ref.~\cite{Madjarov2020}. $A/2\pi$ is $3.958$~GHz~\cite{Atkinson2019} for the $(6s6p)^3P_1$ state, and is $-0.21$~GHz~\cite{Zinkstok2002} for the $(6s6p)^1P_1$ state~(the sign of $A$ is delicate in that some previous measurement suggested a positive $A$~\cite{PhysRev.178.18}). For a highly excited Rydberg electron, its contact with the nucleus is negligible, and the hyperfine interaction of the Rydberg atom is mainly between the nucleus and the inner valence electron, with $A/2\pi\approx12.6$~GHz according to the measurement in Ref.~\cite{PhysRevA.49.3351}. With $n\approx70$, the analysis in Appendix~\ref{appendixB} shows that $A\hat{\mathbf{I}}\cdot \hat{\mathbf{J}}$ couples the $|(6sns)^1S_0,~F=I\rangle$ and $|(6sns)^3S_1~F=I\rangle$ states~(when $F=1/2$) to form two new eigenstates $|\lambda_\pm\rangle$ while $|\lambda_0\rangle=|(6sns)^3S_1~F=I+1\rangle$ remains uncoupled. This leads to three states $\{|\lambda_0\rangle,~|\lambda_+\rangle,~|\lambda_-\rangle\}$ with energies $2\pi\times(6.3,~4.8,~-6.1)$~GHz in reference to the unperturbed energy of the $^{174}$Yb $|(6sns)^1S_0\rangle$ state. For $n\sim70$, Appendix~\ref{appendixB} shows that when we choose $|\lambda_+\rangle$ as $|r\rangle$ a MHz-scale Rydberg Rabi frequency can safely avoid the leakage to other nearby states.

An appropriate intermediate state must have two properties. First, it should possess balanced dipole matrix elements to the ground and Rydberg states. Second, large enough laser power at the transition wavelengths should be available to realize coherent excitations. Several choices are at hand. The $(6s6p)^1P_1$ state, which is $2\pi\times751.5~$THz over the ground state, can have a strong coupling with the ground state, but it has a short lifetime about $5$~ns~\cite{Blagoev1994}. To avoid dissipation from the intermediate state, it should be largely detuned, which makes it challenging to achieve a MHz-scale two-photon Rydberg Rabi frequency. A similar issue exists for $(6s7p)^1P_1$, which is $2\pi\times1216.1~$THz over the ground state and has a short lifetime about $10$~ns. The $(6s6p)^3P_1$ state is $2\pi\times539.4~$THz over the ground state and has a lifetime about $800$~ns~\cite{Blagoev1994}. This makes it a useful choice because a smaller detuning at the intermediate state is enough, so that a MHz-scale two-photon Rabi frequency becomes possible as shown in Appendix~\ref{appendixD}. In principle, the clock state can not be coupled with the ground state, but due to mixing with the $(6s6p)^1P_1^0$ by spin-orbit interaction~\cite{Boyd2007} that results in~(see Appendix~\ref{appendixspinorbit})
\begin{eqnarray}
  |(6s6p)^3P_1\rangle&=&a|(6s6p)^3P_1^0\rangle + b|(6s6p)^1P_1^0\rangle ,\label{eq02}
\end{eqnarray}
it becomes possible to be coupled to the ground state, where the superscript $0$ denotes pure Russell-Saunders coupling and $(a,~b)=(0.991,~-0.133)$~\cite{Budick1970}.

The spin-orbit coupled state in Eq.~(\ref{eq02}) is the key feature that enables our theories. The dipole matrix element between the ground~(Rydberg, with $n=70$) and the intermediate states is on the order of $0.1~(10^{-3})ea_0$, as shown in Appendixes~\ref{appendixB},~\ref{appendixC}, and~\ref{appendixD}. So, fast enough Rabi rotation between the qubit state and $|(6s6p)^3P_1\rangle$ is feasible even though $|b|$ is only $0.133$. On the other hand, $a$ is near 1, and the dipole coupling between $|(6s6p)^3P_1\rangle$ and $|r\rangle$ can reach a large value for $n=70$. The laser wavelengths of about $556$ and $308$~nm for the lower and upper transitions in Fig.~\ref{figure2} are available with commercial or homemade lasers, where the $556$~nm light source has been widely used in spectroscopy~\cite{Pandey2009,Atkinson2019} or atomic trapping and cooling~\cite{Saskin2018,Norcia2018,Lehec2021}. As for the UV lasers~\cite{Wilk2010,Hankin2014}, laser systems with wavelenghs tunable in the range $304-309$~nm~\cite{Higgins2017} were frequently used for exciting Rydberg states of a strontium ion~\cite{Higgins2017,Higgins2017prl,Zhang2020}. Detailed analysis in Appendix~\ref{appendixD} shows that with technically available resources, the effective Rabi frequency between $|1\rangle$ and $|r\rangle$ can be $\Omega_{1r} /2\pi=1.4$~MHz, and meanwhile the Rabi frequency for the transition between $|0\rangle$ and $|r'\rangle=(6s70s)^3S_1~F=3/2,m_F=1/2\rangle$ is $-\Omega_{1r}$.

A $\pi$ pulse with duration $t_\pi$ can complete the transition $|1\rangle\rightarrow-i|r\rangle$. To avoid exciting $|0\rangle$ to the Rydberg state, it is desirable to realize a generalized Rabi frequency $\overline{\Omega} =\sqrt{\Omega_{1r}^2+\Delta^2}$ which is $2N$ times $\Omega_{1r}$, where $N$ is an integer. As can be easily proved~\cite{Shi2017,Shi2018Accuv1}, with a duration $t_\pi$ of the Rydberg lasers, the input state $|0\rangle$ undergoes $N$ detuned Rabi cycles, evolving to
\begin{eqnarray}
 |0\rangle\rightarrow e^{i\alpha}|0\rangle,\label{mappingof0}
\end{eqnarray}
where $\alpha\equiv-[N+\Delta/(2\Omega_{1r}) ]\pi$. Note that to implement the Rydberg gate with Rydberg interaction $V$, there is a phase shift $\beta=\pi|\Omega_{1r}|/(2V)$ to the input state $|11\rangle$ because of the blockade error as studied in Refs.~\cite{Zhang2012,Shi2018prapp2}, extra phase shifts to the laser fields are required to recover a controlled-Z~($C_Z$) gate. More than that, because when the control qubit state $|1\rangle$ is excited to $|r\rangle$, the off-resonant excitation of $|0\rangle$ to $|r'\rangle$ in the target qubit results in a similar phase shift $\beta'=\pi|\Omega_{1r}|/[2(V+\Delta)]$ for the two-qubit input state $|10\rangle$. So, based on the Rydberg blockade gate and following Ref.~\cite{Zhang2012}, we modify the standard $\pi-2\pi-\pi$ pulse sequence of Ref.~\cite{PhysRevLett.85.2208} by replacing the $2\pi$ pulse on the target qubit with two $\pi$ pulses, where the latter $\pi$ pulse has a phase shift $2\alpha$ relative to the the first one. Then, we realize the state transform
\begin{eqnarray}
 |00\rangle&\rightarrow &e^{i4\alpha}|00\rangle,\nonumber\\
 |01\rangle&\rightarrow& -e^{i4\alpha}|01\rangle,\nonumber\\
 |10\rangle&\rightarrow &-e^{i\beta'}|10\rangle,\nonumber\\
 |11\rangle&\rightarrow &-e^{i\beta}|11\rangle.\label{map01}
\end{eqnarray}
Afterwards, single-qubit phase gates can be applied to the control qubit for $|0\rangle\rightarrow e^{-i4\alpha}|0\rangle$ and $|1\rangle\rightarrow e^{-i\beta}|1\rangle$ so that a quasi controlled-Z~($C_Z$) gate is realized with the gate matrix $U=$diag$\{1,~-1,-e^{i(\beta'-\beta)},-1\}$, which differs a little bit from the standard Rydberg $C_Z$ gate $\mathscr{U}=$diag$\{1,~-1,-1,-1\}$. In the standard Rydberg blockade gate~\cite{Saffman2010} studied here, we need $V\gg \Omega_{1r}$, so $\beta'-\beta$ is a tiny number because $\Delta\sim \Omega_{1r}$. Averaging equally over the four possible input states $|00\rangle,~|01\rangle,~|10\rangle$, and $|11\rangle$ and using the fidelity formula in Ref.~\cite{Pedersen2007}, 
\begin{eqnarray}
&& \frac{1}{20}\left[  |\text{Tr}(U^\dag \mathscr{U})|^2 + \text{Tr}(U^\dag \mathscr{U}\mathscr{U}^\dag U ) \right], \label{fidelityError01}
\end{eqnarray}
the error due to different phases accumulated in $|10\rangle$ and $|11\rangle$ is 
\begin{eqnarray}
  E_{\text{ph}} &=& 0.3[1-\cos(\beta'-\beta) ]\nonumber\\
  &\approx&0.15(\beta'-\beta)^2 . \label{phaseerror}
\end{eqnarray}
The error in Eq.~(\ref{phaseerror}) is less than $10^{-5}$ for the parameters shown later. 

Theory 1 can be realized with small magnetic fields. To have, for example, $N=2$, a magnetic field $B\approx3.9$~G~(we use $g_J=g_L$ for brevity; the g factor depends on the choice of $|r\rangle$ shown in Appendix~\ref{appendixB}) can be used to induce a Zeeman shift $\Delta=\sqrt{15}\Omega_{1r}$, so that the generalized Rabi frequency $\overline{\Omega} =\sqrt{\Omega_{1r}^2+\Delta^2}=4\Omega_{1r}$ is realized. The smallness of $B$ is of vital importance since during Rydberg excitation the qubit is released, experiencing free flight and, hence, different magnetic fields. The smaller the magnetic field generated, the smaller its fluctuation.

\subsection{Theory 2}\label{Ybtheory2}
Theory 2 is shown in Fig.~\ref{figure2}(b) where the same symbol $|r\rangle$ is used though $m_F$ is different from that in Fig.~\ref{figure2}(a). The two-photon transition $|1\rangle\rightarrow|r\rangle$ is resonant with right-hand polarized fields. As a consequence there is no way to couple $|0\rangle$ to $(6s70s)^3S_1$ since it does not have a $m_F=5/2$ state. As shown in Appendix~\ref{appendixE}, the angular coupling coefficients for $|1\rangle\rightarrow|r\rangle$ are $\sqrt{3}$ times those in Sec.~\ref{theory1sub}. This basically means that if the field amplitudes are the same for the corresponding lasers used in Fig.~\ref{figure2}(a) and~\ref{figure2}(b), the achievable Rabi frequency for exciting $|1\rangle$ to $|r\rangle$ can be $\Omega_{1r}/2\pi= 2.4$~MHz, leading to a Rydberg-state decay error $7\pi/(4\Omega_{1r})\approx1.1\times10^{-3}$. Because there is a phase shift to $|11\rangle$ due to the imperfect blockade, we follow Ref.~\cite{Zhang2012} to modify the standard $\pi-2\pi-\pi$ pulse sequence of Ref.~\cite{PhysRevLett.85.2208} by replacing the $2\pi$ pulse on the target qubit with two $\pi$ pulses, where the second pulse has a phase shift $\beta$ relative to the the first one. Moreover, as shown later, because the qubit state $|0\rangle$ undergoes highly off-resonant excitation to the intermediate state, there is a certain phase accumulation $\theta$ in $|0\rangle$ for each $\pi$ pulse used for $|1\rangle\leftrightarrow|r\rangle$. Then, Theory 2 can realize the gate 
\begin{eqnarray}
 |00\rangle&\rightarrow &e^{i4\theta}|00\rangle,\nonumber\\
 |01\rangle&\rightarrow& -e^{i(2\theta+\beta)}|01\rangle,\nonumber\\
 |10\rangle&\rightarrow &-e^{i2\theta}|10\rangle,\nonumber\\
 |11\rangle&\rightarrow &-e^{i\beta}|11\rangle,\label{map02}
\end{eqnarray}
where $\theta\approx0.084\pi$ as shown later in Sec.~\ref{numerial} and Fig.~\ref{figure4}(b). Different from Eq.~(\ref{map01}), the state $|10\rangle$ does not have a blockade-induced phase shift because the qubit state $|0\rangle$ can only be excited to the largely detuned intermediate state. Single-qubit phase gates can be applied to both qubits for the transform $|0\rangle\rightarrow e^{-i2\theta}|0\rangle$, and an extra single qubit gate $|1\rightarrow e^{-i\beta}|1\rangle$ can be applied to the target qubit to recover a $C_Z$ gate. The gate by Theory 2 has no phase error as in Eq.~(\ref{phaseerror}). 

Theory 2 differs from Theory 1 in that there is a highly off-resonant excitation from $|0\rangle$ to the intermediate $^3P_1$ state. With the estimate in Appendix~\ref{appendixE}, the ratio between the single-photon Rabi frequency and the detuning to the intermediate state is about $0.01$, which leads to an error less than $5\times10^{-5}$ due to the scattering at $^3P_1$.

 The gates in Eqs.~(\ref{map01}) and~(\ref{map02}) rely on the Rydberg blockade mechanism. To estimate the van der Waals interaction between AEL Rydberg atoms~\cite{Robertson2021}, we use quantum defects in Ref.~\cite{Lehec2018} and radial integration of Ref.~\cite{Kaulakys1995}, and calculate~\cite{Shi2014} $C_6/2\pi=32$~GHz$\mu$m$^6$ for two $(6s70s)^1S_0$ atoms lying along the quantization axis. The interaction for atoms in the $^3S_1$ states should be much larger because they have nine dipole-dipole transition channels for each set of principal quantum numbers, while two atoms in $^1S_0$ only have one, i.e., it couples with $|^1P_1^1P_1\rangle$. But the quantum defects for the $^3P_{0,1,2}$ Rydberg states of ytterbium are not available. To have a conservative estimate~(also because we can choose either $(6s70s)^3S_1$ or $|\lambda_+\rangle$ as $|r\rangle$, see Appendix~\ref{appendixB}), we suppose that the $C_6$ coefficient of $|r\rangle$ is only six times that of $(6s70s)^1S_0$, and, thus, with a qubit spacing $4~\mu$m, the interaction is $V=2\pi\times47$~MHz, which can validate the blockade mechanism with $\Omega_{1r}/2\pi= 1.4~(2.4)$~MHz in Eqs.~(\ref{map01})~[(\ref{map02})] with a blockade error~\cite{Saffman2005}
\begin{eqnarray}
  E_{\text{bl}}=(\Omega_{1r}^2/8)[1/V^2+1/(V+\Delta)^2], \label{blockadeerror}
\end{eqnarray}
for Theory 1, and
\begin{eqnarray}
  E_{\text{bl}}=\Omega_{1r}^2/(8V^2), \label{blockadeerror2}
\end{eqnarray}
for Theory 2. That there is an extra blockade error in Eq.~(\ref{blockadeerror}) is because for Theory 1, the input state $|10\rangle$ also experiences blocked Rydberg excitation, but with a detuning $V+\Delta$. Then, we find that $E_{\text{bl}}$ is about $2.0~(3.3)\times10^{-4}$ for Theory 1~(2).

\begin{figure}
\includegraphics[width=3.2in]
{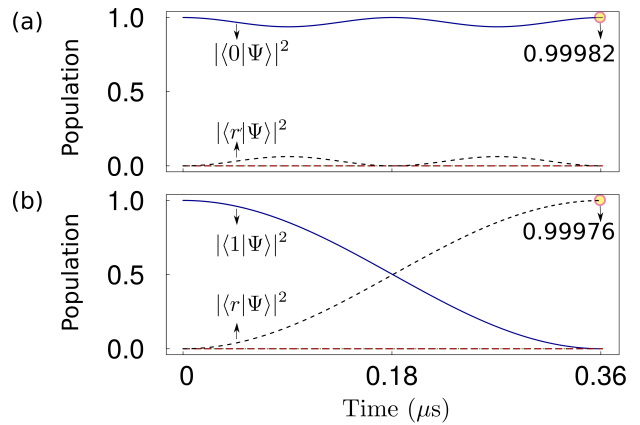}
\caption{Population evolutions by Theory 1 for the input states $|0\rangle$ and $|1\rangle$ in (a) and (b), respectively. The solid~(short dashed) curve denotes population in the qubit~(Rydberg) state, and the dash-dotted and long dashed curves show populations in the two intermediate states $|1(3)/2\rangle$. In both (a) and (b), the populations in $|1(3)/2\rangle$ are on the order of $10^{-4}$ or below during the pulse. The simulations were done with Eq.~(\ref{Hamiltonianfor0}) and Eq.~(\ref{Hamiltonianfor1}), respectively, with parameters specified in the texts near them. The final populations in $|0\rangle$ of (a) and in $|1\rangle$ of (b) are respectively $0.9995$ and $0.9994$ at $t=t_\pi$. But we find that at the moment $t=t_o\equiv0.9994t_\pi$, the populations are $0.99982$ and $0.99976$, respectively. So the pulse duration $0.9994t_\pi$ is better. Also, the phase error is negligible. At $t=t_o$ the value arg$\langle0|\Psi\rangle$ differs from the angle $\alpha$ in Eq.~(\ref{mappingof0})~(with $N=2$) by only $6.0\pi\times10^{-6}$ in (a), and the phase error in (b) is negligible, $1/2-$arg$\langle1|\Psi\rangle/\pi=6.5\times10^{-5}$.  \label{figure3} }
\end{figure}
 
\begin{figure}
\includegraphics[width=3.2in]
{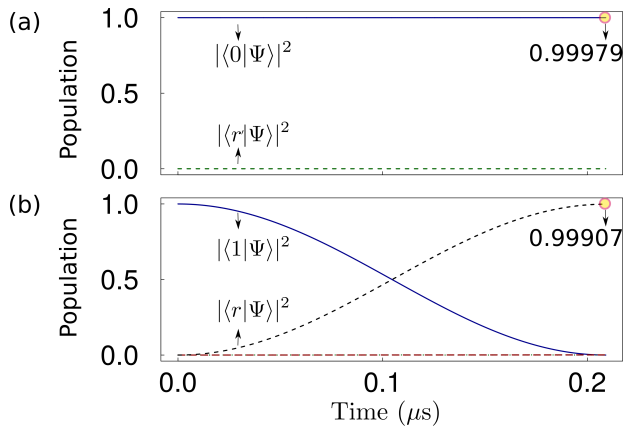}
\caption{Population evolutions by Theory 2 for the input states $|0\rangle$ and $|1\rangle$ in (a) and (b), respectively. The solid~(short dashed) curve denotes population in the qubit~(Rydberg) state, and the dash-dotted and long dashed curves show populations in the two intermediate states $|1(3)/2\rangle$. The simulation in (b) was done with Eq.~(\ref{Hamiltonianfor1}) with parameters specified in the texts near the end of Sec.~\ref{sec03}. The final populations in $|0\rangle$ of (a) and in $|1\rangle$ of (b) are respectively $0.99983$ and $0.99821$ at $t=t_\pi$. But we find that at the moment $t=t_o\equiv0.9994t_\pi$, the populations are $0.99979$ and $0.99907$, respectively. Thus the pulse duration $0.9994t_\pi$ is better as in Fig.~\ref{figure3}. The phase error is negligible in (b), with arg$\langle1|\Psi\rangle/\pi-1/2=8.9\times10^{-5}$ at $t=t_o$. However, there is a phase arg$\langle0|\Psi\rangle/\pi=-0.084$ at $t=t_o$ in (a).  \label{figure4} }
\end{figure}

 \subsection{Numerical test}\label{numerial}
The Hamiltonian for the input state $|1\rangle$ in Theory 1 is
\begin{eqnarray}
  \hat{H}_1&=&   \frac{1}{2} \left(
\begin{array}{cccc}
    2\mathbb{D}_r &\Omega_3^{(1)} &\Omega_1^{(1)}&0\\
     \Omega_3^{(1)} &2\delta_1+2\delta_2 &0&\omega_3^{(1)}\\
    \Omega_1^{(1)}&0 &2\delta_1&\omega_1^{(1)}\\
    0 &\omega_3^{(1)} &\omega_1^{(1)}&2\mathbb{D}_1
    \end{array}
  \right)\label{Hamiltonianfor1}.
\end{eqnarray}
written in the basis $\{|r\rangle,~|3/2\rangle,~|1/2\rangle,~ |1\rangle\}$ where $|r\rangle\equiv|(6s70s)^3S_1,F=3/2,m_F=-1/2\rangle$, and $|3(1)/2\rangle$ are the shorthand notations for the states $|(6s6p)^3P_1,F=3(1)/2,m_F=-1/2\rangle$. Here, $\delta_1$ is the detuning given by the frequency of the lower laser field minus the frequency of the atomic transition and $-\delta_2$ is the hyperfine gap of $^3P_1$. The pulse duration is $t_\pi=\pi/   \Omega_{1r}$, where $\Omega_{1r} = \frac{ \omega_1^{(1)} \Omega_1^{(1)}}{2\delta_1} +\frac{ \omega_3^{(1)} \Omega_3^{(1)}}{2(\delta_1+\delta_2)}$. With experimentally achievable laser field strengths as specified in Appendix~\ref{appendixD}, we can have $( \omega_1^{(1)}, \Omega_1^{(1)},~\delta_1)/2\pi=(-39.9,~103,~2970)$~MHz,~$\delta_2=-2\delta_1$,~$\omega_3^{(1)}=\sqrt{2}\omega_1^{(1)}, \Omega_3^{(1)} =-\Omega_1^{(1)}/\sqrt{2}$. Because the adiabatic elimination of the intermediate states, there will be Stark shifts for the states $|1\rangle$ and $|r\rangle$. To compensate them, we assume that careful calibration is made by, e.g., adding extra highly off-resonant fields~\cite{Maller2015,Shi2020}, to induce effective shifts $\mathbb{D}_1=-(\omega_1^{(1)})^2/(4\delta_1)$ and $\mathbb{D}_r=(\Omega_1^{(1)})^2/(8\delta_1)$. 
 
The Hamiltonian for the input state $|0\rangle$ is
\begin{eqnarray}
  \hat{H}_0&=&   \frac{1}{2} \left(
\begin{array}{cccc}
    2(\mathbb{D}_r+\Delta) &|\Omega_3^{(1)}| &|\Omega_1^{(1)}|&0\\
   |\Omega_3^{(1)}| &2\delta_1+2\delta_2 &0&|\omega_3^{(1)}|\\
    |\Omega_1^{(1)}|&0 &2\delta_1&-|\omega_1^{(1)}|\\
   0 &\omega_3^{(1)} &-|\omega_1^{(1)}|&2\mathbb{D}_1
    \end{array}
  \right),\label{Hamiltonianfor0}
\end{eqnarray}
in the basis $\{|r'\rangle,~|3/2\rangle,~|1/2\rangle,~ |1\rangle\}$. Here, $\Delta=\sqrt{15}\Omega_{1r}$ denotes the Zeeman shift between the states $|r'\rangle$ and $|r\rangle$ from a magnetic field about $3.9$~G; the tiny Zeeman shift between $|0\rangle$ and $|1\rangle$ is absorbed in $\Delta$ in the rotating frame transformation, and the Zeeman shifts in the intermediate states are small compared to the hyperfine gap, and, thus, are neglected. Details for deriving Eqs.~(\ref{Hamiltonianfor1}) and~(\ref{Hamiltonianfor0}) are given in Appendix~\ref{appendixD}.

Figure~\ref{figure3} shows the simulation results with Eqs.~(\ref{Hamiltonianfor1}) and~(\ref{Hamiltonianfor0}). With a pulse duration $t_\pi$, the population transfer is not as accurate as expected because the detunings at the intermediate states $|1(3)/2\rangle$ are finite. But with a pulse duration $t_o=0.9994t_\pi$, the population transfer from $|1\rangle$ to $|r\rangle$ has a fidelity $0.99976$, and meanwhile the population in $|0\rangle$ remains there with a fidelity $0.99982$. After the excitation process shown in Fig.~\ref{figure3}, another pulse of the same duration $t_o$ can be used so that the initial states are restored with fidelities $0.99961$ and $0.99984$ for $|0\rangle$ and $|1\rangle$, respectively. The bizarre fact that the population error in $|1\rangle$ is not twice of the error in the transition $|1\rangle\rightarrow|r\rangle$ results from a peculiar interference between the different transition pathways. The average Rydberg superposition time is $(\pi/4)(7/\Omega_{1r}+3\Omega_{1r}/\overline{\Omega}^2)$, leading to a decay error about $1.9\times10^{-3}$ with $\Omega_{1r}/2\pi=1.4$~MHz and $\tau=330~\mu$s. With the phase error in Eq.~(\ref{phaseerror}), blockade error in Eq.~(\ref{blockadeerror}), intrinsic rotation error, and the Rydberg-state decay, the infidelity of the $C_Z$ gate is $2.4\times10^{-3}$ which is dominated by the Rydberg-state decay. To have a fidelity beyond $0.999$, our results suggest that to host qubits in cryogenic chambers is necessary. 

In Theory 2, the Hamiltonian for the input state $|1\rangle$ is still Eq.~(\ref{Hamiltonianfor1}) but with different parameters because circularly polarized fields are used in Fig~(\ref{figure2}(b) as detailed in Appendix~\ref{appendixE}. With similar strengths of laser fields used in the simulation for Fig.~\ref{figure3}, we have $( \omega_1^{(1)}, \Omega_1^{(1)},~\delta_1)/2\pi=(56.4,~126,~2970)$~MHz,~$\delta_2=-2\delta_1$,~$\omega_3^{(1)}=\omega_1^{(1)}/\sqrt{2}, \Omega_3^{(1)} =-\sqrt{2}\Omega_1^{(1)}$. The required extra detunings at the ground and Rydberg states are $\mathbb{D}_1=(\omega_1^{(1)})^2/(8\delta_1)$ and $\mathbb{D}_r=-(\Omega_1^{(1)})^2/(4\delta_1)$, respectively. For the input state $|0\rangle$, it is excited to $|3/2\rangle$ with detuning $\delta_1+\delta_2$ and Rabi frequency $\omega_3^{(1)}=2\pi\times49$~MHz. By these parameters, the population dynamics for the two input states $|0(1)\rangle$ are shown in Fig.~\ref{figure4}. As in Fig.~\ref{figure3}, we find that a pulse duration $t_o=0.9994t_\pi$, instead of $t_\pi$, can lead to a more accurate Rydberg excitation. In sharp contrast to Theory 1, the phase accumulation in the input state $|0\rangle$ can not be neglected. At the end of the pulse we find arg$\langle0|\Psi\rangle/\pi=-0.084$. This phase is accumulated due to the detuned Rabi oscillation between the ground state and the intermediate state $|3/2\rangle$ with a generalized Rabi frequency $\overline{\omega}\equiv\sqrt{(\omega_3^{(1)})^2+(\delta_1+\delta_2)^2}$~\cite{Shi2017,Shi2018Accuv1}. The phase is accumulated in a way similar to that in Eq.~(\ref{mappingof0}), but with a much larger $\overline{\omega}$. In each generalized Rabi cycle, there is a phase accumulation $-\pi[1+(\delta_1+\delta_2)/\overline{\omega}]$, leading to $\pi(\delta_1+\delta_2+\overline{\omega})/|\Omega_{1r}|\approx-0.084\pi$ for a pulse duration $\pi/|\Omega_{1r}|$. This extra phase in $|0\rangle$ can be effectively eliminated by the single-qubit phase gates as shown below Eq.~(\ref{map02}). So, including the blockade error in Eq.~(\ref{blockadeerror2}), the intrinsic rotation error, and the Rydberg-state decay, the infidelity of the $C_Z$ gate by Theory 2 is $2.0\times10^{-3}$ that is again dominated by the Rydberg-state decay.

\begin{figure}
\includegraphics[width=3.2in]
{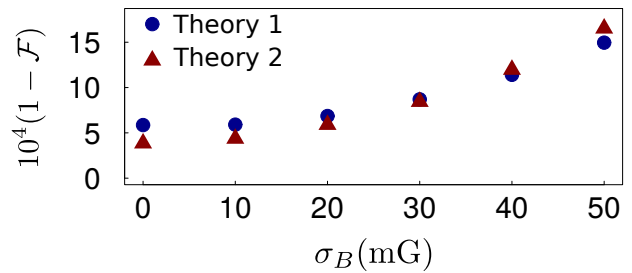}
\caption{Gate error~(scaled by $10^4$) due to the fluctuation of magnetic field $\sigma_B$ in units of mG~(=$10^{-7}$~T). Round~(rectangular) symbols denote results for Theory 1~(2). The numerical evaluation assumes perfect Rydberg blockade via Eqs.~(\ref{Hamiltonianfor1})-(\ref{eq14}) with the gate fidelity defined in Eq.~(\ref{fidelityError01}).   \label{figure-MagError} }
\end{figure}
\subsection{Gate error due to the fluctuation of magnetic fields}\label{secD}
The fluctuation of magnetic fields can cause frequency jitter in the atomic transitions between the ground, intermediate, and Rydberg states. Both the two qubit states have zero electron angular momentum, and the g factor is simply that of the nuclear spin. Because the nuclear magneton is about 1836 times smaller than the Bohr magneton, the level shifts in the qubit states can be neglected when estimating the gate error from the field fluctuation.  

When the magnetic field is exactly equal to the value we desire, the Hamiltonians for the states $|1\rangle$ and $|0\rangle$ are described by Eqs.~(\ref{Hamiltonianfor1}) and~(\ref{Hamiltonianfor0}), respectively. With a fluctuation $\sigma_B$ of the magnetic field, Eqs.~(\ref{Hamiltonianfor1}) and~(\ref{Hamiltonianfor0}) become
\begin{eqnarray}
 \text{Eq.}(\ref{Hamiltonianfor1}) -\text{diag}\{g_R,~g_{i},~g_{i},~g_g \}\sigma_B \mu_B/2,
\end{eqnarray}
and
\begin{eqnarray}
 \text{Eq.}(\ref{Hamiltonianfor0}) +\text{diag}\{g_R,~g_{i},~g_{i},~g_g \}\sigma_B \mu_B/2,\label{eq14}
\end{eqnarray}
respectively, where the subscripts $R,~i,$ and $g$ distinguish the Rydberg, intermediate, and ground states, and the minus and plus signs in the two lines above originate from the two different Zeeman levels associated with the two qubit states. As discussed above, we set $g_g=0$ here since the nuclear spin Zeeman splitting is tiny compared to those of the electrons. The Rydberg g factor $g_R$ depends on the state we choose, shown in Appendix~\ref{appendixB}. For brevity, we take $g_R=1$ as an example. The g factors for the two intermediate states should be calculated according to the mixing between the singlet and triplet states shown in Eq.~(\ref{eq02}), $g_p=a^2g(^{3}P_1)+b^2g(^1P_1)\approx1.49$~\cite{PhysRev.128.1159,Aymar1984,PhysRevA.49.4443}, where $g(^{3}P_1) $ and $g(^1P_1)$ are $1.5$ and $1$, respectively. 

To quantify the error from the magnetic field fluctuation, we use Eq.~(\ref{fidelityError01}) by defining $\mathscr{U}$ as the ideal gate matrix, and calculate $U$ for each nonzero $\sigma_B$. In the Rydberg blockade gate~\cite{PhysRevLett.85.2208,Saffman2010}, the blockade error is estimated by averaging gate errors with interactions varying around the desired $V$. This is because the blockade error can accidentally vanish~\cite{Shi2020jpb} for a certain $V$ that depends on the impossible condition of absolutely static atoms. Thus, we use Eq.~(\ref{blockadeerror}) for a useful estimate of the blockade error but assume large enough $V$ here to estimate the error solely from $\sigma_B$. 

Numerical results with $\sigma_B$ up to $50$~mG are shown in Fig.~\ref{figure-MagError}, where the round~(rectangular) symbols denote results for Theory 1~(2). One can see that when no field fluctuation occurs, the rotation error is about $5\times10^{-4}$. This is because as shown in Figs.~\ref{figure3} and~\ref{figure4}, there are intrinsic rotation errors in the two-photon Rydberg excitations. Figure~\ref{figure-MagError} shows that with a field fluctuation around $10$~mG, the rotation error is nearly the same as when there is no field fluctuation. As shown in the Supplemental Materials of Ref.~\cite{Graham2019}, $\sigma_B<10$~mG is achievable with $B=6$~G that is of similar magnitude as used here. Even with a large fluctuation $\sigma_B=50$~mG, Fig.~\ref{figure-MagError} shows that our method enables a fidelity over 0.996 for both Theory 1 and Theory 2 when including the phase error in Eq.~(\ref{phaseerror}), the blockade error in Eq.~(\ref{blockadeerror}), and the Rydberg-state decay.

 \section{ Rydberg excitation of nuclear spin qubits with $^{87}$Sr}\label{sec04}
 It is useful to extend our theories to the case of $^{87}$Sr which was widely studied in experiments~\cite{Ido2003,Ye2013,Gaul2016,Winchester2017,Cooper2018,Covey2019,Norcia2018,Ding2018,Teixeira2020,Madjarov2020} and theories~\cite{Werij1992,Boyd2007,Vaillant2012,Vaillant2014,Dunning2016,Robicheaux2019,Mukherjee2011,Robertson2021}. Because of the different quantum defects in ytterbium and strontium~\cite{Vaillant2012}, we calculate that the $C_6$ coefficient of strontium is 20 times that of ytterbium for a certain $^1S_0$ Rydberg state. Thus it is possible to set a larger qubit spacing if strontium atoms are used for Rydberg blockade so as to have less crosstalk in the Rydberg lasers. Indeed, the Rydberg blockade interaction of both cold and hot strontium atoms have been experimentally studied~\cite{Zhang2015,DeSalvo2016,Yoshida2017,Madjarov2020}, demonstrating the applicability of strontium in Rydberg atom quantum science.
 
 The key to the shown theories in Sec.~\ref{sec03} is the hyperfine interactions and spin-orbit couplings. However, such interactions are strikingly different for different AEL atoms. For $^{87}$Sr with $I=9/2$, the qubit states to encode information can be chosen from the states $m_I=7/2$ and $9/2$, whose degeneracy can be split by a field of the order of $1$~G that leads to a kHz-scale Zeeman shift with a nuclear magnetic moment $-1.0936\mu_N$~\cite{Sansonetti2010}. To coherently excite the nuclear spin states to Rydberg states, however, we find that the lowest $^3P_1$ state is no longer the best choice for the intermediate state. This is because for strontium, Eq.~(\ref{eq02}) becomes
 \begin{eqnarray}
  |(5s5p)^3P_1\rangle&=&a|(5s5p)^3P_1^0\rangle +b|(5s5p)^1P_1^0\rangle ,\label{eq02strontium}
 \end{eqnarray}
 with $b=-0.0286$~\cite{Boyd2007}, which is about five times smaller than that for ytterbium. This leads to a rather small Rabi frequency if laser fields of similar magnitudes to those in Sec.~\ref{sec03} are used.

One solution to the above issue is to use the state $|(5s6p)^1P_1\rangle$ as the intermediate state which can be coupled to the $^1S_0$ Rydberg states. As detailed in Appendix~\ref{appendixF}, $|(5s6p)^1P_1\rangle$ has a linewidth $\Gamma=2\pi\times0.3$~MHz and it can be coupled to the ground state with a dipole matrix element that is about half of $|\langle  (6s6p)^3P_1  || er|| (6s^2) ^1S_0\rangle|$ of ytterbium. If we choose the $|(5s70s)^1S_0\rangle$ state as $|r\rangle$ to entangle the qubits, the achievable Rabi frequencies for $|1\rangle\rightarrow|r\rangle$ are $0.8$ times those in Fig.~\ref{figure2} if the same values $\delta_{1(2)}~\sim2\pi\times3$~GHz are used. In principle, larger Rydberg Rabi frequencies can be achieved by using smaller detunings. More details are shown in Appendix~\ref{appendixF}. Finally, we note that although the $^1S_0$ Rydberg states do not have large Zeeman shifts, they are coupled to the $^3S_1$ Rydberg states via the hyperfine interaction. Such a coupling induces a large energy separation on the order of the hyperfine constant, so that we can choose one state as $|r\rangle$. The hyperfine induced state mixing was studied in detail in Ref.~\cite{Ding2018}; see also Appendix~\ref{appendixB}. In this case, a small magnetic field can lead to a MHz-scale $\Delta$ by the electron Zeeman shift although there can be a weak diamagnetic effect in the $^1S_0$ state~\cite{Fonck1977}. 

We calculate that the van der Waals interaction coefficient for the strontium $(5s70s)^1S_0$ state is $C_6/2\pi=-710$~GHz$\mu$m$^6$ with the quantum defects used in Ref.~\cite{Robicheaux2019,Vaillant2012}. Note that if stronger interactions are desired, one can choose the state near $n=63$ where F\"{o}rster resonance occurs~\cite{Robicheaux2019}. With a qubit spacing of $5~\mu$m the interaction is $|V|/2\pi=45$~MHz, which is large enough to validate the blockade gate in the form of Eq.~(\ref{map02}). Compared to the ytterbium case studied at the end of Sec.~\ref{Ybtheory2}, here the qubit spacing is $1~\mu$m larger while the value of $V$ is almost the same. So, using $^{87}$Sr is more favorable to avoid crosstalk compared to $^{171}$Yb.

\section{Discussion}\label{sec05}
We have studied exciting nuclear spin qubit states to Rydberg states for entanglement generation. As for single-qubit gates, their implementation with nuclear spins in AEL atoms has been studied in, e.g., Refs.~\cite{Stock2008,Daley2008,Pagano2019}. It is also an open question whether using Rydberg excitation as introduced in this work can we have fast and accurate single-qubit gates. However, a straightforward extension of Theory 1 to single-qubit gate operation is at hand. In the clock transition from the ground $^1S_0$ state to the lowest $^3P_0$ state, the transition is possible because there is a small mixing of the $^1P_0^0$ component in the $^3P_0$ state. A detailed analysis in Ref.~\cite{Boyd2007} shows that with a magnetic field applied there will be a differential shift for the transition $|^1S_0,m_I\rangle\rightarrow|^3P_0,m_I\rangle$ considering different $m_I$. For the case of $^{87}$Sr, the shift is 110$\times m_I$Hz$/$G. This basically means that by applying an appropriate B-field we can excite the two nuclear spin qubit states from the ground state to the $^3P_0$ state, one resonantly with a kHz-scale Rabi frequency $\Omega_0$ while the other with a generalized Rabi frequency $\overline{\Omega}$ because it is detuned. The detuning can be specifically set so that $\overline{\Omega}$ is $N$ times $\Omega$, where $N$ is an integer. Then, it is possible to induce single-qubit gates based on the exotic selective excitation of nuclear spin qubits as in Theory 1.  

The gates in Eq.~(\ref{map01}) for Theory 1 and Eq.~(\ref{map02}) for Theory 2 belong to the standard Rydberg blockade gate introduced in Ref.~\cite{PhysRevLett.85.2208}. Based on the Rydberg blockade mechanism, there have been various protocols. For example, simultaneous Rydberg excitation of both qubits can lead to entangling gates in the blockade regime. A special advantage of Theory 1 is that both qubit states experience Rydberg excitation, and different $\Delta$ can be easily obtained by different B-fields. It is an open question whether by choosing appropriate ratio between $\Delta$ and the Rydberg Rabi frequencies can we generate a $C_Z$ gate with one step. This should be possible since for alkali-metal atoms, fast $C_Z$-like gates~\cite{Levine2019} or CNOT gates~\cite{Shi2020} can be generated with effective one-pulse sequences.

We have based our discussions on the Rydberg blockade mechanism. It is also of interest to extend our theories to Rydberg gates by the antiblockade method~\cite{Ates2007,Amthor2010} as reviewed in Ref.~\cite{Su2020epl}. Although the antiblockade regime is sensitive to the fluctuation of the Rydberg interactions, the simultaneous excitation of both qubit states to Rydberg state, as shown in Theory 1, can be useful for designing exotic logic gates~\cite{Shi2018Accuv1}. With short enough Rydberg superposition times, the error from fluctuation of interactions can be small.

\section{Conclusions}\label{sec06}
We provide two theories to realize Rydberg excitation of one of the two nuclear spin ground states of an AEL atom. We perform a detailed study with ytterbium and briefly study strontium when using our theories. The first theory needs an external magnetic field on the order of 1~G to work, while the second method can work also with larger magnetic fields. We have shown methods to realize $C_Z$ gates based on these two theories and found that fidelities over $0.997$ are achievable for both methods with feasible resources. The key advantage of our theories is that strong magnetic fields are not required so that fast dephasing from the noises in the magnetic field can be avoided in practical experiments. Because AEL atoms like ytterbium can be easily cooled, can be cooled without altering the information encoded in the nuclear spin space, and their nuclear spins are insensitive to magnetic fluctuations in the environment, our theories can lead to new opportunities for Rydberg atom quantum science and technology.

\section*{ACKNOWLEDGMENTS}
The author thanks Yan Lu for insightful discussions. This work is supported by the National Natural Science Foundation of China under Grants No. 12074300 and No. 11805146, the Natural Science Basic Research plan in Shaanxi Province of China under Grant No. 2020JM-189, and the Fundamental Research Funds for the Central Universities.

\appendix{}

\section{Spin-orbit coupling between the low-lying $^1P_1^0$ and $^3P_1^0$ states }\label{appendixspinorbit}
The theory can be easily tested in experiments because the quantum control over alkaline-earth-metal atoms and ytterbium was well studied~\cite{Yamamoto2016,Saskin2018,Cooper2018,Covey2019,Madjarov2020,Norcia2018}. Here, we give extra details that can be useful to set up laser fields for the excitation scheme shown in the main text. To be specific, we consider ytterbium-171. 

There is a spin-orbit coupling between the $|(6s6p)^1P_1^0\rangle$ and $|(6s6p)^3P_1^0\rangle$ states, where the superscript $0$ denotes pure Russell-Saunders states. The spin-orbit coupling leads to~\cite{Boyd2007}
\begin{eqnarray}
  |(6s6p)^1P_1\rangle&=& a|(6s6p)^1P_1^0\rangle - b|(6s6p)^3P_1^0\rangle,\nonumber\\
  |(6s6p)^3P_1\rangle&=& b|(6s6p)^1P_1^0\rangle + a|(6s6p)^3P_1^0\rangle,\label{spinorbit01}
\end{eqnarray}
where $(a,~b)$ depend on the atomic species. For ytterbium-171 we have $(a,~b)=(0.991,~-0.133)$ according to Ref.~\cite{Budick1970}. For another ytterbium isotope with nuclear spin, i.e., $^{173}$Yb, a similar mixing occurs, with $(a,~b)=(0.990,~-0.141)$~\cite{PhysRev.128.1159}.

\section{Eigenstates with hyperfine splitting}\label{appendixB}
In this appendix we study the hyperfine induced state mixing between the singlet and triplet $s$-orbital Rydberg states. The hyperfine coupling
\begin{eqnarray}
 \hat{V}_{\text{HI}} &=& A\hat{\mathbf{I}}\cdot \hat{\mathbf{J}}
\end{eqnarray}
induces energy shifts~\cite{Lurio1962,Ding2018}
\begin{eqnarray}
0&=&  \langle (6sns)^1S_0,F=I|\hat{V}_{\text{HI}}|(6sns)^1S_0,F=I\rangle ,\nonumber\\
-A/2&=& \langle (6sns)^3S_1,F=I|\hat{V}_{\text{HI}}|(6sns)^3S_1,F=I\rangle ,\nonumber\\
A/2&=&  \langle (6sns)^3S_1,F=I+1|\hat{V}_{\text{HI}}|(6sns)^3S_1,\nonumber\\
&& F=I+1\rangle , \label{eqB2}
\end{eqnarray}
for all cases of AEL atoms, and 
\begin{eqnarray}
  -A(I+1)/2&=& \langle (6sns)^3S_1,F=I-1|\hat{V}_{\text{HI}}|(6sns)^3S_1,\nonumber\\
  &&F=I-1\rangle ,
\end{eqnarray}
for AEL atoms with $I>1$. 

There is also state mixing between $|(6sns)^1S_0,F=I\rangle$ and $|(6sns)^3S_1,F=I\rangle$. To be specific, we consider $^{171}$Yb and label the two states by $|s1\rangle\equiv|(6sns)^1S_0,F=1/2\rangle$ and $|s2\rangle\equiv|(6sns)^3S_1,F=1/2\rangle$. For illustration, we consider principal quantum numbers around 70 and study consequences from different $n$ at the end of this section. The singlet-triplet gap $1.8\times10^6/n^{\ast3}$~GHz between $|s1\rangle$ and $|s2\rangle$ is about  $E_s\approx2\pi\times5$~GHz with an effective principal quantum number $n^{\ast}=70$. To understand the hyperfine interaction in the state mixing between the two states $|s1\rangle$ and $|s2\rangle$, we follow the method in Refs.~\cite{Lehec2018,Ding2018}. With the energy of $^{174}$Yb as a reference, the Hamiltonian for $^{171}$Yb is
\begin{eqnarray}
\hat{H}(171) &=& \hat{H}_0(174,m_{171})+\hat{V}_{\text{HI}},
\end{eqnarray}
where $\hat{H}_0(174,m_{171})$ is the Hamiltonian of $^{174}$Yb rescaled by the isotope shift. Here $m_{171}=m_eM_{171}/(m_e+M_{171})$, where $m_e$ is the mass of an electron and $M_{171}$ is the mass of $^{171}$Yb$^+$. In a high-lying Rydberg state, the hyperfine interaction is dominated by the interaction between the inner $6s$ valence electron and the nuclear spin because the highly excited electron is far from the nucleus and, thus, has negligible contact with the nucleus. We take the energy of $^{174}$Yb from the eigenenergy of $\hat{H}_0(174,m_{171})$ and treat the hyperfine interaction $\hat{V}_{\text{HI}}$ as a perturbation. In the basis of $\{|s2\rangle,~|s1\rangle\}$, $\hat{V}_{\text{HI}}$ is given by~\cite{Lurio1962}
\begin{eqnarray}
  \left(
  \begin{array}{cc}
    E_s -A/2& \sqrt{3}A/4\\
     \sqrt{3}A/4 &0
    \end{array}
  \right),\label{equationHI}
\end{eqnarray}
where $A$ is the nuclear magnetic dipole constant and $E_s\approx 2\pi\times  5$~GHz with $n\sim 70$. According to the measurement in Ref.~\cite{PhysRevA.49.3351}, $A/2\pi\approx12.6$~GHz. Equation~(\ref{equationHI}) has two eigenstates, 
\begin{eqnarray}
 |\lambda_\pm\rangle& \propto& [D\mp (E_s -A/2) ]|s1\rangle +\sqrt{3}A |s2\rangle/2,\label{lambda+-}
\end{eqnarray}
where 
\begin{eqnarray}
 D &=& \sqrt{ 3A^2/4+ (E_s -A/2)^2 },
\end{eqnarray}
and the eigenenergies are
\begin{eqnarray}
 \lambda_\pm &=& (E_s/2 -A/4)\pm D/2,\label{lambdapm}
\end{eqnarray}
which are equal to $2\pi\times4.8$ and $-6.1$~GHz, respectively. Then, due to the hyperfine interaction, the energies for the three eigenstates $|(6sns)^3S_1,~F=3/2\rangle$,~$|\lambda_+\rangle$, and~$|\lambda_-\rangle$ are $2\pi\times(6.3,~4.8,~-6.1)$~GHz, respectively, where $2\pi\times6.3$~MHz is from Eq.~(\ref{eqB2}). For the purpose in this work, one can choose $|r\rangle=|(6sns)^3S_1,~F=3/2\rangle$ because there will be more states that can be coupled by the dipole-dipole interaction for $|^3S_1~^3S_1\rangle$ compared to the state $|^1S_0~^1S_0\rangle$, leading to larger van der Waals interactions~\cite{Vaillant2012,Robicheaux2019}. For the purpose of state detection it is usually useful to couple Rydberg states to the rapidly decaying low-lying state $(6s6p)^1P_1$. However, with this choice the coupling can be largely limited by the small triplet component in the wavefunction of $(6s6p)^1P_1$ as shown in Eq.~(\ref{spinorbit01}).

Another choice is to choose $|\lambda_+\rangle$ as $|r\rangle$. Then, $|\lambda_+\rangle$ is separated from $|(6sns)^3S_1,~F=3/2\rangle$ and~$|\lambda_-\rangle$ by $2\pi\times1.5$ and $2\pi\times10.9$~GHz, respectively. $|\lambda_+\rangle$ has an amplitude $\langle s2|\lambda_+\rangle\approx -0.66$ in the addressable state $|s2\rangle$~(while $\langle s1|\lambda_+\rangle\approx 0.75$), which should be considered when estimating the Rabi frequencies. Because the separation between $|r\rangle$ and the other two states is more than $2\pi\times1.5$~GHz, we can ignore their excitation for a MHz-scale $\Omega$. There is also a mixing between $|(6sns)^1S_0,F=1/2\rangle$ and $|(6s(n\pm 1)s)^3S_1,F=1/2\rangle$ with a much smaller coupling constant $\sqrt{3}A/40$~\cite{Ding2018}. For $n\sim70$ considered in this paper, the energy separation between $(6s(n\pm 1)s)^3S_1$ and $(6sns)^3S_1$ is about $2\pi\times24$~GHz~(estimated from the quantum defects of the $^1S_0$ states~\cite{Lehec2018}), which implies a portion less than $10^{-3}$ for the states with principal quantum numbers $(n\pm 1)$ in $|\lambda_\pm\rangle$. So we can use the description in Eq.~(\ref{equationHI}).

Note that because $|(6s6p)^3P_1\rangle$ consists of both singlet and triplet states, there is some chance to excite the $|(6s6p)^1P_1^0\rangle$ state to the $|(6snd)^1D_2\rangle$ state. But according to the measured results in~\cite{Lehec2018}, the energy of $(6s70s)^1S_0$ is $50417.67$cm$^{-1}$, and the two $^1D_2$ states nearest to it are $(6s68d)^1D_2$ and $(6s69d)^1D_2$ with energies $50417.33$ and $50418.10$cm$^{-1}$, respectively. This means that the gap to the nearest two $^1D_2$ states from $(6s70s)^1S_0$ is $2\pi\times10.2$~GHz. Besides, there is some possibility to couple $|(6s6p)^3P_1^0,F=1(3)/2\rangle$ to the $|(6snd)^3D_{1,2,3}\rangle$ states. But according to the measured results in~\cite{Lehec2018}, the energy of $(6s70s)^1S_0$ is $50417.67$cm$^{-1}$, and the two $^1D_2$ states nearest to it are $(6s68d)^3D_2$ and $(6s69d)^1D_2$ with energies $50417.30$ and $50418.07$cm$^{-1}$, respectively; converting to Hz, these two states are away from $(6s70s)^1S_0$ by $2\pi\times(10.2,~12.0)$~GHz. Because the fine gap in high-lying Rydberg states is negligible with $n=70$, these data show that the state $|r\rangle$ is far separated from nearby states that may be addressed from $(6s6p)^3P_1$.

As a conclusion, we can choose either $|(6sns)^3S_1,~F=3/2\rangle$ or $|\lambda_+\rangle$ as $|r\rangle$. For $n=70$, the gap to nearby Rydberg states is large if a MHz-scale Rydberg Rabi frequency is employed so that the leakage can be safely neglected. If $|\lambda_+\rangle$ is chosen as $|r\rangle$, the mixing between singlet and triplet states in it means that the Land\'{e} factor $g_R$ for the Rydberg state differs from the g factor of the $^3S_1^0$ Rydberg state. With the data shown above, we have $g_R=2 |\langle s2|\lambda_+\rangle|^2\approx0.87$. This g factor is more than two times smaller than that of $|(6sns)^3S_1,~F=3/2\rangle$. In order to show numerical examples that can catch the phenomena for both choices of the Rydberg state, we take a value in between with, for example, $g_R=1$.  

\section{Lifetimes and dipole matrix elements}\label{appendixC}

The lifetimes of high-lying Rydberg states of strontium are not wellknown but can be estimated from measurement results of low-lying Rydberg states. A search in the literature shows that the highest ytterbium Rydberg $^1S_0$ state whose lifetime was ever measured has a principal quantum number $n=26$~\cite{Fang2001}. The measured lifetimes for the ytterbium $^1S_0$ Rydberg states with principal quantum numbers $n\in[21,~26]$ show that their lifetimes are mainly limited by the blackbody radiation at room temperatures~\cite{Fang2001}. Moreover, Ref.~\cite{Fang2001} shows that at room temperatures the lifetime is well approximated by $\tau\approx1.17(n-\mu)^3$~ns for these Rydberg states. According to the recent measurement in~\cite{Lehec2018}, we have $\mu\approx4.278$. By this, we can estimate that the lifetime for the $(6sns)^1S_0$ state is $332~\mu$s for $n=70$. 

The achievable Rabi frequency is determined by the dipole matrix elements of the relevant transitions. Because two-photon Rabi frequencies are used, we shall consider both the lower and the upper transitions. In particular, we compare two possible choices with $\pi$ polarized laser fields~(where the $r_0$ component in the spherical basis is used in the estimate of the dipole matrix elements).

\subsection{$(6s6p)^1P_1$ as the intermediate state}
The first choice is $(6s^2)~^1S_0\rightarrow (6s6p)^1P_1\rightarrow (6sns)^1S_0$ with two reduced dipole matrix elements $\mathcal{D}_l$ and $\mathcal{D}_u$, which are the ${}^{2S+1}L_J$-dependent reduced dipole matrix elements for the lower and upper transitions, respectively,
\begin{eqnarray}
  \mathcal{D}_l&=& -\langle (6s6p)^1P_1^0 || er|| (6s6s)^1S_0^0\rangle,\nonumber\\
  \mathcal{D}_u&=& -\langle  (6sns)^1S_0^0 || er|| (6s6p)^1P_1^0\rangle, \label{dipoledefinition01}
\end{eqnarray}
where $e$ is the elementary charge and the superscript $0$ denotes states formed via the ideal Russell-Saunders coupling~(otherwise there can be singlet-triplet mixing). 

Here, we choose the transition $(6s^2)^1S_0\rightarrow (6s6p)^1P_1\rightarrow (6sns)^1S_0$. In principle, we can also consider $(6s^2)^1S_0\rightarrow (6s6p)^1P_1\rightarrow (6sns)^3S_1$ since $(6s6p)^1P_1\rightarrow (6sns)^3S_1$ is possible because  the state $(6s6p)^1P_1$ has a $(6s6p)^3P_1^0$ component with a coefficient $-b$ shown in Eq.~(\ref{spinorbit01}). However, the dipole coupling to the Rydberg state is already very tiny, and the factor $-b$ is small.

\subsection{$(6s6p)^3P_1$ as the intermediate state}
The second choice is $(6s^2)^1S_0\rightarrow (6s6p)^3P_1\rightarrow (6sns)^3S_1$. However, the transition between singlet and triplet states is spin forbidden. But as shown in Eq.~(\ref{spinorbit01}), the state $(6s6p)^3P_1$ has a $(6s6p)^1P_1^0$ component with a coefficient $b$. The reduced dipole matrix element for the lower transition is still
\begin{eqnarray}
  \mathcal{D}_l&=& -\langle (6s6p)^1P_1^0 || er|| (6s6s)^1S_0\rangle,\label{dipoledefinition02}
\end{eqnarray}
while the upper one is no longer
\begin{eqnarray}
  \mathcal{D}_u&=& -\langle  (6sns)^1S_0^0 || er|| (6s6p)^1P_1^0\rangle, \nonumber
\end{eqnarray}
but becomes
\begin{eqnarray}
  \mathbb{D}_u&=& -\langle  (6sns)^3S_1 || er|| (6s6p)^3P_1^0\rangle, \label{dipoledefinition03}
\end{eqnarray}
if $| (6s6p)^3S_1\rangle$ is used. Here there is no singlet-triplet state mixing for the considered Rydberg states with $F=I+1$ as shown around Eq.~(\ref{equationHI}), so we do not use the superscript ``0'' for the Rydberg state $|r\rangle$. To compare $ \mathcal{D}_u$ and $ \mathbb{D}_u$, we use angular coupling rules to further reduce the radial couplings to 
\begin{eqnarray}
  \mathcal{D}_u&=& -\sqrt{3}\left\{\begin{array}{ccc}0&1 &1\\1&0&0\end{array}
  \right\}\langle  (6sns)S || er|| (6s6p) P\rangle\nonumber\\
  &=&-\langle  (6sns)S || er|| (6s6p) P\rangle ,\nonumber\\
  \mathbb{D}_u&=& \sqrt{3}\left\{\begin{array}{ccc}0&1 &1\\1&1&1\end{array}
  \right\}\langle  (6sns)S || er|| (6s6p) P\rangle\nonumber\\
  &=&-\frac{1}{\sqrt{3}}\langle  (6sns)S || er|| (6s6p) P\rangle ,\label{furtherreduction}
\end{eqnarray}
where $\{\cdots\}$ is a 6-j symbol and the integration $\langle  (6sns)S || er|| (6s6p) P\rangle$ is about $-0.0046ea_0$ for $n=70$~\cite{Covey2019pra} and, thus, $\mathbb{D}_u\approx-1.5\times10^{-3}ea_0$.

We then examine the dipole matrix element $\langle  (6s6p)^1P_1 || er|| (6s^2) ^1S_0\rangle$ for the lower transition $(6s^2)~^1S_0\rightarrow (6s6p)^1P_1$ which has a linewidth $\Gamma=2\pi\times29$~MHz~\cite{Yamamoto2016,Saskin2018}. As can be verified~\cite{DASteck} by using the Weisskopf-Wigner approximation, the dipole matrix element can be estimated by the following relation~[see Eq.~(11.33) of Ref.~\cite{DASteck}]
\begin{eqnarray}
 \Gamma &=& \frac{\omega_0^3}{\pi \epsilon_0\hbar c^3} |\langle  (6s6p)^1P_1 || er|| (6s^2) ^1S_0\rangle|^2,\label{Weisskopf}
\end{eqnarray}
 where $\epsilon_0$ is the free-space dielectric permittivity, $c$ is the light speed in vacuum, $\hbar$ is the Planck constant, and $\omega_0/2\pi\approx7.5\times10^{14}$~Hz is the transition frequency, which lead to $|\langle  (6s6p)^1P_1  || er|| (6s^2) ^1S_0\rangle|=1.38ea_0$. Note that the above estimate has not taken into account of the singlet-triplet mixing as shown in Eq.~(\ref{spinorbit01}). In other words, the actual coupling matrix element should include the factor $a$ in Eq.~(\ref{spinorbit01}). However, $a$ is near 1 and, hence, we have $|\langle  (6s6p)^1P_1^0  || er|| (6s^2) ^1S_0\rangle|=1.38ea_0$.

 The dipole matrix element for the upper transition can be approximated by the semiclassical analytical formulas~\cite{Kaulakys1995} which have been tested in~\cite{Walker2008}. This method is applicable here since the Rydberg state we consider is relatively high, and we can safely assume that the Rydberg electron is far away from the core formed by the nuclear spin and the other electrons in the atom. We have $|\langle  (6s70s)S || er|| (6s6p) P\rangle|=0.0046ea_0$ with the estimate in Ref.~\cite{Covey2019pra}. However, there is a singlet-triplet state mixing in the Rydberg state, which should be considered if either $|\lambda_+\rangle$ or $|\lambda_-\rangle$ of Eq.~(\ref{lambda+-}) is used as $|r\rangle$.%

The above data show that the choice of $(6s6p)^3P_1$ as the intermediate state is more favorable because of Eq.~(\ref{spinorbit01}), where $(6s6p)^3P_1$ mainly has the $(6s6p)^3P_1^0$ component and thus can be easily excited to the $(6sns)^3S_1$ state with a dipole matrix element $a\mathbb{D}_u\approx\mathbb{D}_u$. Because of the large transition dipole matrix element between the ground state and the $(6s6p)^1P_1^0$ component in $(6s6p)^3P_1$, the lower transition still has a sizable dipole matrix element $b\mathcal{D}_l$. 

\begin{figure}
\includegraphics[width=3.2in]
{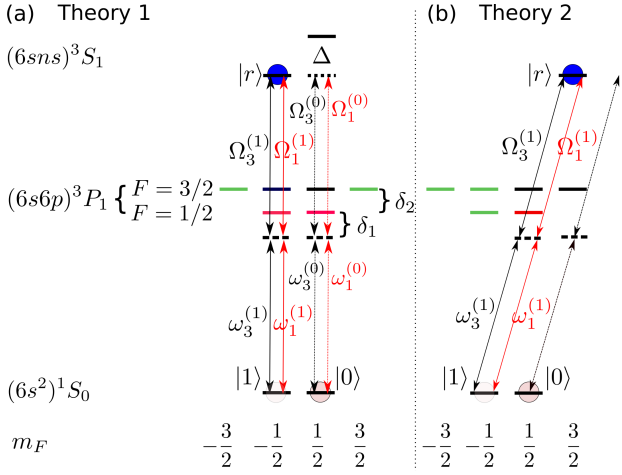}
 \caption{Rydberg excitation of $|1\rangle$ with Theory 1 in (a), and with Theory 2 in (b). Both the $|0\rangle$ and $|1\rangle$ states of the qubit are excited. In (a), the four Rabi frequencies for exciting the qubit state $|0\rangle$ is indicated, where the red arrows indicate the transition via the $F=3/2$ sublevel of the intermediate state, while the black arrows are for transitions via the $F=1/2$ sublevel of the intermediate state. In both (a) and (b), the four Rabi frequencies $\omega_3^{(1)}, \omega_1^{(1)},  \Omega_3^{(1)}$, and $\Omega_1^{(1)}$ are indicated for exciting the transition from $|1\rangle$ to $|r\rangle$.   }\label{Rabi-appendix}
\end{figure}

\section{Rabi frequencies in Theory 1}\label{appendixD}
Here, we study Rabi frequencies by choosing the pure triplet Rydberg state as $|r\rangle$ for brevity; data with $|\lambda_+\rangle$ as $|r\rangle$ can be obtained by slightly modifying the results. The laser fields are linearly polarized along the quantization axis for Theory 1. The two qubit states $|0\rangle$ and $|1\rangle$ are both excited to Rydberg states, shown in Fig.~\ref{Rabi-appendix}(a). The used Rydberg state is then $|r\rangle$=$|(6sns)^3S_1,F=3/2,m_F=-1/2\rangle$ that can be excited from $|1\rangle\equiv|^1S_0,m_I=-1/2\rangle$. From the ground state to the intermediate state, the $(6s6p)^3P_1^0$ component in $(6s6p)^3P_1$ is not coupled because the transition between triplet and singlet states is spin forbidden. But the $(6s6p)^1P_1^0$ component in $(6s6p)^3P_1$ can be coupled. For the upper transition to the Rydberg state $(6sns)^3S_1$, only the $(6s6p)^3P_1^0$ component in $(6s6p)^3P_1$ is coupled. In essence, the transition is like
\begin{eqnarray}
  &&  |1\rangle\rightarrow|(6s6p)^1P_1^0,F=\frac{1(3)}{2},m_F=-\frac{1}{2}\rangle~[\text{in}~(6s6p)^3P_1] \nonumber\\
  && |(6s6p)^3P_1^0,F=\frac{1(3)}{2},m_F=-\frac{1}{2}\rangle~[\text{in}~(6s6p)^3P_1] \rightarrow|r\rangle,\nonumber\\
\end{eqnarray}
for $|1\rangle$, and similarly for $|0\rangle$ because the two qubit states are symmetrical to each other. In other words, their transitions to Rydberg states are characterized with similar dipole matrix elements and transition selection rules. To show this, one can calculate the four one-photon Rabi frequencies for the transition from $|1\rangle$ to $|r\rangle$,
\begin{eqnarray}
  \omega_3^{(1)} &=& \mathcal{E}_l \mathcal{D}_lC_3^{(1)} , 
 \nonumber\\ 
  \omega_1^{(1)} &=& \mathcal{E}_l \mathcal{D}_lC_1^{(1)}  ,
 \nonumber\\
  \Omega_3^{(1)} &=& \mathcal{E}_u \mathbb{D}_uC_3^{(1)'} , 
 \nonumber\\ 
  \Omega_1^{(1)} &=& \mathcal{E}_u \mathbb{D}_uC_1^{(1)'} ,\label{appC01}
\end{eqnarray}
where the dipole matrix elements are given by Eqs.~(\ref{dipoledefinition02}) and~(\ref{dipoledefinition03}), and the coefficients $C_3^{(1)},~C_1^{(1)},~C_3^{(1)'}$, and $C_1^{(1)'}$ are determined by the dipole selection rules and the Wigner-Eckart theorem,
\begin{eqnarray}
  C_3^{(1)}  &=&-\sqrt{6}\left\{\begin{array}{ccc}1&0 &1\\\frac{1}{2}&\frac{3}{2}&\frac{1}{2}\end{array}
  \right\}
  \mathbb{C}_{\frac{\overline{1}}{2}0 \frac{\overline{1}}{2}}^{\frac{1}{2}1 \frac{3}{2}}\nonumber\\
  &=&-\sqrt{6}\cdot(- \sqrt{1/6})\cdot\sqrt{2/3} ,\nonumber
\end{eqnarray}
\begin{eqnarray}
  C_1^{(1)}  &=&-\sqrt{6}\left\{\begin{array}{ccc}1&0 &1\\\frac{1}{2}&\frac{1}{2}&\frac{1}{2}\end{array}
  \right\}
  \mathbb{C}_{\frac{\overline{1}}{2}0 \frac{\overline{1}}{2}}^{\frac{1}{2}1 \frac{1}{2}}\nonumber\\
  &=&-\sqrt{6}\cdot\sqrt{1/6}\cdot(-\sqrt{1/3})  ,\nonumber
\end{eqnarray}
\begin{eqnarray}
  C_3^{(1)'}  &=&2\sqrt{3}\left\{\begin{array}{ccc}1&1 &1\\\frac{3}{2}&\frac{3}{2}&\frac{1}{2}\end{array}
  \right\}
  \mathbb{C}_{\frac{\overline{1}}{2}0 \frac{\overline{1}}{2}}^{\frac{3}{2}1 \frac{3}{2}}\nonumber\\
  &=&2\sqrt{3}\cdot(\sqrt{10}/12)\cdot(-\sqrt{1/15}) ,\nonumber
\end{eqnarray}
\begin{eqnarray}
  C_1^{(1)'}  &=&-\sqrt{6}\left\{\begin{array}{ccc}1&1 &1\\\frac{1}{2}&\frac{3}{2}&\frac{1}{2}\end{array}
  \right\}
  \mathbb{C}_{\frac{\overline{1}}{2}0 \frac{\overline{1}}{2}}^{\frac{1}{2}1 \frac{3}{2}}\nonumber\\
  &=&-\sqrt{6}\cdot (-1/6) \cdot(\sqrt{2/3})  , \label{CGappC01}
\end{eqnarray}
where $\mathbb{C}_{\cdots}^{\cdots}$ denotes a Clebsch-Gordan coefficient and $\{\cdots\}$ is a 6-j symbol. From the above results we can derive by adiabatic elimination~\cite{Shi2014} that the effective two-photon Rabi frequency between $|1\rangle$ and $|(6sns)^3S_1,F=3/2,m_F=-1/2\rangle$ is 
\begin{eqnarray}
  \Omega_{1r} &=& -\frac{ \omega_1^{(1)} \Omega_1^{(1)}}{2\delta_1} -\frac{ \omega_3^{(1)} \Omega_3^{(1)}}{2(\delta_1+\delta_2)}\nonumber\\
  &=&- \frac{  \mathcal{E}_l \mathcal{E}_u \mathcal{D}_l\mathbb{D}_u}{3\sqrt{3}} \left[\frac{ 1}{2\delta_1}-\frac{ 1}{2(\delta_1+\delta_2)} \right]\label{rabi1r}
\end{eqnarray}
where $\delta_1$ is given by the frequency of the lower laser field minus the frequency of the atomic transition $|1\rangle\leftrightarrow|(6sns)^3P_1,F=1/2,m_F=-1/2\rangle$, and $-\delta_2$ is the hyperfine gap between the $F=1/2$ and $F=3/2$ levels. We have $-\delta_2=2\pi\times5.94$~GHz according to~\cite{Pandey2009,Atkinson2019}. By deriving the effective Rabi frequency, there is some stark shift which we assume to be compensated by calibrating the detunings, field amplitudes, and if necessary, extra highly off-resonant dressing fields~\cite{Maller2015,Shi2020}. In numerical simulations these shifts are included, shown in Eqs.~(\ref{Hamiltonianfor1}) and~(\ref{Hamiltonianfor0}).

To verify that the transitions from $|0\rangle$ to the Rydberg state has similar property as that from $|1\rangle$, we study the four coefficients related with the Rabi frequencies for the transition from $|0\rangle$, 
\begin{eqnarray}
C_3^{(0)}  &=&-\sqrt{6}\left\{\begin{array}{ccc}1&0 &1\\\frac{1}{2}&\frac{3}{2}&\frac{1}{2}\end{array}
  \right\}
  \mathbb{C}_{\frac{1}{2}0 \frac{1}{2}}^{\frac{1}{2}1 \frac{3}{2}}\nonumber\\
  &=&-\sqrt{6}\cdot(- \sqrt{1/6})\cdot\sqrt{2/3} ,\nonumber
\end{eqnarray}
\begin{eqnarray}
  C_1^{(0)}  &=&-\sqrt{6}\left\{\begin{array}{ccc}1&0 &1\\\frac{1}{2}&\frac{1}{2}&\frac{1}{2}\end{array}
  \right\}
  \mathbb{C}_{\frac{1}{2}0 \frac{1}{2}}^{\frac{1}{2}1 \frac{1}{2}}\nonumber\\
  &=&-\sqrt{6}\cdot\sqrt{1/6}\cdot\sqrt{1/3}  ,\nonumber
\end{eqnarray}
\begin{eqnarray}
  C_3^{(0)'}  &=&2\sqrt{3}\left\{\begin{array}{ccc}1&1 &1\\\frac{3}{2}&\frac{3}{2}&\frac{1}{2}\end{array}
  \right\}
  \mathbb{C}_{\frac{1}{2}0 \frac{1}{2}}^{\frac{3}{2}1 \frac{3}{2}}\nonumber\\
  &=&2\sqrt{3}\cdot(\sqrt{10}/12)\cdot \sqrt{1/15} ,\nonumber
\end{eqnarray}
\begin{eqnarray}
  C_1^{(0)'}  &=&-\sqrt{6}\left\{\begin{array}{ccc}1&1 &1\\\frac{1}{2}&\frac{3}{2}&\frac{1}{2}\end{array}
  \right\}
  \mathbb{C}_{\frac{1}{2}0 \frac{1}{2}}^{\frac{1}{2}1 \frac{3}{2}}\nonumber\\
  &=&-\sqrt{6}\cdot (-1/6) \cdot(\sqrt{2/3})  , \label{CGappC01}
\end{eqnarray}
from which one can derive the effective Rabi frequency from $|0\rangle$ to the Rydberg state $|(6sns)^3S_1,F=3/2,m_F=1/2\rangle$ as 
\begin{eqnarray}
  \Omega_{0r} &=& -\frac{ \omega_1^{(0)} \Omega_1^{(0)}}{2\delta_1} -\frac{ \omega_3^{(0)} \Omega_3^{(0)}}{2(\delta_1+\delta_2)}\nonumber\\
  &=&\frac{  \mathcal{E}_l \mathcal{E}_u \mathcal{D}_l\mathbb{D}_u}{3\sqrt{3}} \left[\frac{ 1}{2\delta_1}-\frac{ 1}{2(\delta_1+\delta_2)} \right].\label{rabi0r}
\end{eqnarray}

In order to have a larger Rabi frequency, a positive $\delta_1$ that is smaller than $-\delta_2$ is preferred. The transition from the ground state to the $(6s6p)^3P_1$ state needs light of wavelength of about $556$~nm~\cite{Pandey2009,Atkinson2019}, and its transition to the Rydberg state with $n=70$ needs radiation of wavelength $308.4$~nm~(According to Ref.~\cite{Lehec2018} the energy of $6s70s$ state is $50417.7$cm$^{-1}$, while the energy of $(6s6p)^3P_1$ state is $17992.01$cm$^{-1}$~\cite{Blagoev1994}). Although the $308$~nm UV laser appeared less frequently in the literature, a laser system with wavelengh tunable in the range $304-309$~nm was used in the experiments~\cite{Higgins2017} to excite Rydberg states of a strontium ion~(see also Refs.~\cite{Higgins2017prl,Zhang2020} where similar lasers were used for Rydberg excitations of ions). Such UV lasers were usually prepared by frequency doubling via second-harmonic generation~(see the supplementary information of Ref.~\cite{Zhang2020}) where similar powers can be obtained for similar frequencies, and, thus, we analyze achievable laser powers by the more frequently used 319~nm lasers.

The lower transition has a dipole matrix element on the order of $ea_0$ as shown below Eq.~(\ref{Weisskopf}). Such a magnitude is of similar magnitude to that of the lower transition in exciting a Rydberg state of rubidium or cesium where Rabi frequencies up to $2\pi\times300$~MHz could be achieved~\cite{Wilk2010}, and, hence, we assume $\omega_1^{(1)}/2\pi=300$~MHz. However, the dipole matrix element for the upper transition is tiny, and then the Rabi frequency for the upper transition is mainly determined by the achievable strength of the electric field $\mathcal{E}_u$ in the UV lasers. In Ref.~\cite{DeSalvo2016}, Rydberg excitation from the $(5s5p)^3P_1$ of strontium was achieved with ultraviolet light at $319$~nm with a power 34mW. In Ref.~\cite{Zhang2020}, a $306$~nm laser with power 60~mW focused to a waist of $8~\mu$m was used for preparing Rydberg ions~(Ref.~\cite{Zhang2020} indicates that their laser power can in principle be increased by two orders of magnitude). In Ref.~\cite{Hankin2014}, Rydberg excitation of the $84p$ state of cesium was achieved by a 319~nm light with a laser power 300mW. Thus, we assume a laser power 300mW in our estimate. To avoid error in the laser Rabi frequencies from position fluctuation of atoms, we assume a beam waist $10~\mu$m as in the experiment in~\cite{Hankin2014}. Then $\mathcal{E}_u=\sqrt{2I/(c\epsilon_0)}\approx2.5\times10^6V/m$, where $I\approx 300mW/(100\pi\mu$m$^2$), $c$ is the speed of light in vacuum, and $\epsilon_0$ is the free space permittivity. This is a strong field but it is experimentally feasible; see, e.g., page 2 of Ref.~\cite{Hankin2014} that shows a similarly huge UV laser field with intensity $6\times10^8$W$/m^2$ at the atoms. With $|\mathbb{D}_u|\approx0.0046ea_0/3$ for $n=70$~\cite{Covey2019pra}~[here the factor $1/3$ is from Eq.~(\ref{furtherreduction})], we have $\Omega_1^{(1)} =  \mathcal{E}_u \mathcal{D}_uC_1^{(1)'}/\hbar\approx2\pi\times103$~MHz. From Eq.~(\ref{CGappC01}) we have $\omega_1^{(1)}\Omega_1^{(1)} = -\omega_1^{(0)} \Omega_1^{(0)}$, and, thus, setting $\delta_1=-(\delta_1+\delta_2)$ can maximize the effective Rabi frequency, which requires $\delta_1=-\delta_2/2=2\pi\times2.97$~GHz. The above estimate shows that reaching a Rydberg Rabi frequency $ |\Omega_{1r}| /2\pi = 300\cdot 103/ (2970)\approx10.4$~MHz is possible. However, according to Eq.~(\ref{spinorbit01}), Appendix~\ref{appendixB}, and Appendix~\ref{appendixC}, the actual Rabi frequency for the lower transition shall be multiplied by a factor of $b$, and, thus, the final achievable Rabi frequency is 
\begin{eqnarray}
  |\Omega_{1r}| /2\pi& =& (300b)\cdot 103/ 2970\nonumber\\
  &\approx&1.4\text{ MHz}.\label{Rabifinal}
\end{eqnarray}
Note that a large ratio between $\delta_{1(2)}$ and the single-photon Rabi frequencies is required in deriving the effective two-photon Rabi frequency. In the above case, the ratios are $2970/(300b)\approx74$ and $2970/ 103\approx29$ for the lower and upper transitions, respectively. This can lead to high-fidelity control over the Rydberg excitations. 

Comparing Eq.~(\ref{rabi0r}) and Eq.~(\ref{rabi1r}) shows that the state $|0\rangle$ is excited to Rydberg states with a Rabi frequency $- \Omega_{1r}$. Note that the magnetic field can induce some energy difference between the intermediate states for exciting $|0\rangle$ and $|1\rangle$. But compared to the GHz-scale $\delta_{1(2)}$, they are negligible when the applied magnetic field is on the order of 1~G that does not change the picture described above. In a magnetic field the splitting between $|0\rangle$ and $|1\rangle$ is $ g_{I} \mu_nB\approx2\pi\times1.5$~kHz with $B=3.9$~G and $\mu_{n}=0.4919\mu_N$ for $^{171}$Yb~\cite{Porsev2004}. The $m_F=\pm1/2$ levels of the Rydberg states have a Zeeman shift $g_{ J} \mu_BB$. In this work we assume $g_J=g_L=1$ although it is $0.87$ for the choice of $|r\rangle$ analyzed below Eq.~(\ref{lambdapm}). 

\section{Rabi frequencies in Theory 2}\label{appendixE}
In Theory 2, circularly polarized laser fields are used, as shown in Fig.~\ref{Rabi-appendix}(b). Because the $^3S_1$ state does not have a $m_F=5/2$ state, the qubit state $|0\rangle$ can not be excited to the Rydberg state, and only $|1\rangle$ can go to the Rydberg state. Both the lower and upper fields are right-hand polarized, with field amplitudes $\mathcal{E}_l$ and $\mathcal{E}_u$, respectively. We calculate the four one-photon Rabi frequencies for the transition from $|1\rangle$ to $|r\rangle$,
\begin{eqnarray}
  \omega_3^{(1)} &=& \mathcal{E}_l \mathcal{D}_lC_3^{(1)} , 
 \nonumber\\ 
  \omega_1^{(1)} &=& \mathcal{E}_l \mathcal{D}_lC_1^{(1)}  ,
 \nonumber\\
  \Omega_3^{(1)} &=& \mathcal{E}_u \mathbb{D}_uC_3^{(1)'} , 
 \nonumber\\ 
  \Omega_1^{(1)} &=& \mathcal{E}_u \mathbb{D}_uC_1^{(1)'} ,\label{theory2appC01}
\end{eqnarray}
where the dipole matrix elements are given by Eqs.~(\ref{dipoledefinition02}) and~(\ref{dipoledefinition03}), and the coefficients $C_3^{(1)},~C_1^{(1)},~C_3^{(1)'}$, and $C_1^{(1)'}$ are determined by the dipole selection rules and the Wigner-Eckart theorem,
\begin{eqnarray}
  C_3^{(1)}  &=&-\sqrt{6}\left\{\begin{array}{ccc}1&0 &1\\\frac{1}{2}&\frac{3}{2}&\frac{1}{2}\end{array}
  \right\}
  \mathbb{C}_{\frac{\overline{1}}{2}1 \frac{1}{2}}^{\frac{1}{2}1 \frac{3}{2}}\nonumber\\
  &=&-\sqrt{6}\cdot(- \sqrt{1/6})\cdot\sqrt{1/3} ,\nonumber
\end{eqnarray}
\begin{eqnarray}
  C_1^{(1)}  &=&-\sqrt{6}\left\{\begin{array}{ccc}1&0 &1\\\frac{1}{2}&\frac{1}{2}&\frac{1}{2}\end{array}
  \right\}
  \mathbb{C}_{\frac{\overline{1}}{2}1 \frac{1}{2}}^{\frac{1}{2}1 \frac{1}{2}}\nonumber\\
  &=&-\sqrt{6}\cdot\sqrt{1/6}\cdot(-\sqrt{2/3})  ,\nonumber
\end{eqnarray}
\begin{eqnarray}
  C_3^{(1)'}  &=&2\sqrt{3}\left\{\begin{array}{ccc}1&1 &1\\\frac{3}{2}&\frac{3}{2}&\frac{1}{2}\end{array}
  \right\}
  \mathbb{C}_{\frac{1}{2}1 \frac{3}{2}}^{\frac{3}{2}1 \frac{3}{2}}\nonumber\\
  &=&2\sqrt{3}\cdot(\sqrt{10}/12)\cdot(-\sqrt{2/5}) ,\nonumber
\end{eqnarray}
\begin{eqnarray}
  C_1^{(1)'}  &=&-\sqrt{6}\left\{\begin{array}{ccc}1&1 &1\\\frac{1}{2}&\frac{3}{2}&\frac{1}{2}\end{array}
  \right\}
  \mathbb{C}_{\frac{1}{2}1 \frac{3}{2}}^{\frac{1}{2}1 \frac{3}{2}}\nonumber\\
  &=&-\sqrt{6}\cdot (-1/6) \cdot(1)  . \label{CGappC01theory2}
\end{eqnarray}
 Comparing Eq.~(\ref{CGappC01}) and Eq.~(\ref{CGappC01theory2}) shows that the Rabi frequencies $(\omega_3^{(1)},  \omega_1^{(1)},  \Omega_3^{(1)} ,  \Omega_1^{(1)} )$ here are $(1/\sqrt{2},\sqrt{2},\sqrt{6},\sqrt{6}/2)$ times those in Eq.~(\ref{appC01}). From the above results we can derive by adiabatic elimination~\cite{Shi2014} that the effective two-photon Rabi frequency between $|1\rangle$ and $|(6sns)^3S_1,F=3/2,m_F=-1/2\rangle$ is 
\begin{eqnarray}
  \Omega_{1r} &=&- \frac{ \omega_1^{(1)} \Omega_1^{(1)}}{2\delta_1} -\frac{ \omega_3^{(1)} \Omega_3^{(1)}}{2(\delta_1+\delta_2)}\nonumber\\
  &=&- \frac{  \mathcal{E}_l \mathcal{E}_u \mathcal{D}_l\mathbb{D}_u}{3} \left[\frac{ 1}{2\delta_1}-\frac{ 1}{2(\delta_1+\delta_2)} \right], \label{rabi1rtheory2}
\end{eqnarray}
which is $\sqrt{3}$ times Eq.~(\ref{rabi1r}) mainly because the angular factor is larger than that in deriving Eq.~(\ref{rabi1r}) because of different polarizations in the laser fields. With a similar analysis leading to Eq.~(\ref{Rabifinal}), we estimate that the Rydberg Rabi frequency for $|1\rangle\rightarrow|r\rangle$ can reach $2\pi\times2.4$~MHz. For the transition from $|0\rangle$ to $|(6s6p)^3P_1,F=3/2,m_F=3/2\rangle$ with a detuning $\delta_1+\delta_2\approx2\pi\times(-2.97)$~GHz, the Rabi frequency is
\begin{eqnarray}
  \omega_3^{(0)} &=& \mathcal{E}_l \mathcal{D}_lC_3^{(0)} , 
 \end{eqnarray}
with
\begin{eqnarray}
  C_3^{(0)}  &=&-\sqrt{6}\left\{\begin{array}{ccc}1&0 &1\\\frac{1}{2}&\frac{3}{2}&\frac{1}{2}\end{array}
  \right\}
  \mathbb{C}_{\frac{1}{2}1 \frac{3}{2}}^{\frac{1}{2}1 \frac{3}{2}}\nonumber\\
  &=&-\sqrt{6}\cdot(- \sqrt{1/6})\cdot1 ,\nonumber
\end{eqnarray}
which leads to $300\sqrt{3/2}\beta\approx2\pi\times49$~MHz according to the estimate above Eq.~(\ref{Rabifinal}). This means that the leakage to the intermediate state is on the order of $(40/2970)^2/2\approx1.4\times10^{-4}$ that is negligible compared to the decay errors. On the other hand, the tiny population on the $^3P_1$ state also leads to decay from it. However, this is negligible. For a $\pi$ pulse on the qubit state $|1\rangle$, the time for the atom initialized in the qubit state $|0\rangle$ to stay at the $^3P_1$ level is about $(49/2970)^2\cdot\pi/|2\Omega_{1r}|\approx0.028$~ns, which leads to an extra decay error $3.5\times10^{-5}$ from $^3P_1$ by taking a lifetime $800$~ns for it~\cite{Blagoev1994}. Such an error is orders of magnitude smaller than that from the Rydberg-state decay and, hence, can be neglected.

\section{Data with $^{87}$Sr}\label{appendixF}
The spin-orbit coupling in $^{87}$Sr leads to
\begin{eqnarray}
  |(5s5p)^1P_1\rangle&=& a|(5s5p)^1P_1^0\rangle - b|(5s5p)^3P_1^0\rangle,\nonumber\\
  |(5s5p)^3P_1\rangle&=& b|(5s5p)^1P_1^0\rangle + a|(5s5p)^3P_1^0\rangle,\label{spinorbit01sr}
\end{eqnarray}
where $(a,~b)$ depend on the strength of the spin-orbit coupling in the specific AEL element. We have $(a,~b)=(0.9996,~-0.0286)$ for strontium according to Ref.~\cite{Boyd2007}. Compared to Eq.~(\ref{spinorbit01}) for ytterbium, the mixing characterized by $b$ is about five times weaker here, which leads to a long lifetime $21.5~\mu$s~\cite{Zelevinsky2006} of strontium $|(5s5p)^3P_1\rangle$~(the lifetime of ytterbium $|(5s5p)^3P_1\rangle$ is less than $1~\mu$s).

For the dipole matrix element $\langle  (5s5p)^1P_1 || er|| (5s^2) ^1S_0\rangle$ with a linewidth of $\Gamma=2\pi\times32$~MHz~\cite{Millen2010,Mukherjee2011} and transition frequency $\omega_0/2\pi\approx6.5\times10^{14}$~Hz, we use 
\begin{eqnarray}
 \Gamma &=& \frac{\omega_0^3}{\pi \epsilon_0\hbar c^3} |\langle  (5s5p)^1P_1 || er|| (5s^2) ^1S_0\rangle|^2,\label{Weisskopf02}
\end{eqnarray}
to estimate $|\langle  (5s5p)^1P_1  || er|| (5s^2) ^1S_0\rangle|=1.80ea_0$. This means that if we choose $|(5s5p)^3P_1\rangle$ as the intermediate state, the dipole matrix element between it and the ground state is only $|1.80ea_0\cdot b|\approx0.05ea_0$. Compared to the case with ytterbium where $|\langle  (6s6p)^3P_1  || er|| (6s^2) ^1S_0\rangle|=0.18ea_0$, Rydberg excitation via the $|(5s5p)^3P_1\rangle$ state of strontium is much more difficult.

A more useful choice for the intermediate state is $|(5s6p)^1P_1\rangle$ with a linewidth $\Gamma=2\pi\times0.3$~MHz and a $293$~nm excitation wavelength~(which can be prepared via second-harmonic generation~\cite{Higgins2017,Higgins2017prl,Zhang2020}). With a method similar to Eq.~(\ref{Weisskopf02}), we find that $|\langle  (5s6p)^1P_1  || er|| (5s^2) ^1S_0\rangle|=0.088ea_0$. This means that the $|(5s6p)^1P_1\rangle$ state is indeed a useful choice. But the coupling between the intermediate and Rydberg $^3S_1$ states is spin forbidden, so we consider the $^1S_0$ Rydberg state. However, the $^1S_0$ Rydberg state is strongly mixed with the $^3S_1$ Rydberg state~\cite{Ding2018}. As studied in Ref.~\cite{Ding2018}, the mixing leads to two split states with a GHz-scale energy separation, so it is possible to use the hyperfine-mixed state as $|r\rangle$. Importantly, because of the triplet component in $|r\rangle$ there can also be a strong Zeeman shift when a $B$-field on the order of 1~G is applied.

With $|(5s6p)^1P_1\rangle$ as the intermediate state, the dipole matrix element $\mathcal{D}_u$ for the upper transition here is of similar magnitude to that for the upper transition of ytterbium in Fig.~\ref{Rabi-appendix}. This is because the final reduced matrix elements are both expressed as $\langle  ns  || er|| 6p\rangle$. According to Eq.~(\ref{furtherreduction}), however, $|\mathcal{D}_u|$ is $\sqrt{3}$ times larger than $|\mathbb{D}_u|$, which means that the matrix element between the $|(5s6p)^1P_1\rangle$ and $|(5s70s)^1S_0\rangle$ states of $^{87}$Sr is $\sqrt{3}$ times larger than that of Fig.~\ref{Rabi-appendix}. Then, the achievable Rydberg Rabi frequency for $|1\rangle\rightarrow |70^1S_0\rangle$ is $\sqrt{3}\cdot0.088ea_0/(0.18ea_0)\approx0.8$ times those in Eqs.~(\ref{Rabifinal}) and~(\ref{rabi1rtheory2}) for Theory 1 and Theory 2, respectively. Of course, such an estimate assumes the same hyperfine gap $|\delta_2|$ and detuning $\delta_{1}$ as in Fig.~\ref{Rabi-appendix}. In practice, this is not crucial since we can choose appropriate $\delta_1$ so as to have a large enough Rydberg Rabi frequency described above; the negative effect is that with smaller $\delta_1$ there can be more scattering from the intermediate state. 

%

\end{document}